\documentclass[12pt]{spieman}  
\usepackage{amsmath,amsfonts,amssymb}
\usepackage{graphicx}
\usepackage{setspace}
\usepackage{tocloft}

\usepackage{txfonts}
\usepackage{upgreek}
\usepackage{acronym}
\usepackage{array}
\usepackage{xcolor}
\usepackage{lineno}

\acrodef{ADI}[ADI]{Angular Differential Imaging}
\acrodef{AO}[AO]{adaptive optics}
\acrodef{ExAO}[ExAO]{extreme adaptive optics}
\acrodef{AR}[AR]{anti-reflection}
\acrodef{DM}[DM]{deformable mirror}
\acrodef{DPSS}[DPSS]{diode-pumped solid-state }
\acrodef{FLICE}[FLICE]{femtosecond laser inscribed chemical etching}
\acrodef{FWHM}[FWHM]{full-width at half-maximum}
\acrodef{HRS}[HRS]{high resolution spectroscopy}
\acrodef{IFS}[IFS]{integral field spectrograph}
\acrodef{IFU}[IFU]{Integral Field Unit}
\acrodef{LQG}[LQG]{Linear Quadratic Gaussian}
\acrodef{LSF}[LSF]{line spread function}
\acrodef{MagAO-X}[MagAO-X]{Magellan Adaptive Optics Extreme }
\acrodef{MCF}[MCF]{multi-core fiber}
\acrodef{MCIFU}[MCIFU]{multi-core fiber-fed integral field unit}
\acrodef{MFD}[MFD]{mode field diameter}
\acrodef{MLA}[MLA]{microlens array}
\acrodef{MMF}[MMF]{multi-mode fiber}
\acrodef{NA}[NA]{numerical aperture}
\acrodef{NCPA}[NCPA]{non common path abberation}
\acrodef{NIR}[NIR]{near infra-red}
\acrodef{NYRIA}[NYRIA]{Network of Young Researchers in Instrumentation for Astronomy}
\acrodef{SM}[SM]{single-mode}
\acrodef{SMF}[SMF]{single-mode fiber}
\acrodef{SDI}[SDI]{Spectral Differential Imaging}
\acrodef{RDI}[RDI]{Reference star Differential Imaging}
\acrodef{PA}[PA]{path abberation}
\acrodef{PGMEA}[PGMEA]{propylene-glycol-methyl-ether-acetate}
\acrodef{PSF}[PSF]{point spread function}
\acrodef{QSS}[QSS]{quasi-static speckles}
\acrodef{RCWA}[RCWA]{Rigorous Coupled Wave Analysis}
\acrodef{rms}[rms]{root-mean square}
\acrodef{SCAO}[SCAO]{single-conjugate adaptive optics}
\acrodef{SNR}[SNR]{signal-to-noise ratio}
\acrodef{TNA}[TNA]{transnational access}
\acrodef{ULI}[ULI]{ultrafast laser inscription}
\acrodef{VPHG}[VPHG]{Volume Phase Holographic Grating}
\acrodef{WHT}[WHT]{William Herschel Telescope}

\newcommand{\remove}[1]{}
\newcommand*{\doi}[1]{\href{http://dx.doi.org/#1}{doi: #1}}

\title{Diffraction-limited integral-field spectroscopy for extreme adaptive optics systems with the Multi-Core fiber-fed Integral-Field Unit}

\author[a,b,*]{Sebastiaan Y. Haffert}
\author[c,l]{Robert J. Harris}
\author[d]{Alessio Zanutta}
\author[e]{Fraser A. Pike}
\author[d]{Andrea Bianco}
\author[d]{Eduardo Redaelli}
\author[e]{Aur\'elien Beno\^{\i}t}
\author[e]{David G. MacLachlan}
\author[e]{Calum A. Ross}
\author[f,m]{Itandehui Gris-S\'anchez}
\author[g,h]{Mareike D. Trappen}
\author[g,h]{Yilin Xu}
\author[g,h]{Matthias Blaicher}
\author[g,h]{Pascal Maier}
\author[d]{Giulio Riva}
\author[i]{Baptiste Sinquin}
\author[i]{Caroline Kulcs\'{a}r}
\author[j]{Nazim Ali Bharmal}
\author[k]{Eric Gendron}
\author[j]{Lazar Staykov}
\author[j]{Tim J. Morris}
\author[l]{Santiago Barboza}
\author[l]{Norbert Muench}
\author[j]{Lisa Bardou}
\author[i]{L\'eonard Preng\`ere}
\author[i]{Henri-Fran\c cois Raynaud}
\author[c]{Phillip Hottinger}
\author[c,n]{Theodoros Anagnos}
\author[j]{James Osborn}
\author[g,h]{Christian Koos}
\author[e]{Robert R. Thomson}
\author[f]{Tim A. Birks}
\author[a]{Ignas A. G. Snellen}
\author[a]{Christoph U. Keller}
\affil[a]{ Leiden Observatory, Leiden University, PO Box 9513, Niels Bohrweg 2, 2300 RA Leiden, The Netherlands}
\affil[b]{Steward Observatory, Unversity of Arizona, 933 North Cherry Avenue, Tucson, Arizona}
\affil[c]{Zentrum f\"ur Astronomie der Universit\"at Heidelberg, Landessternwarte K\"onigstuhl, K\"onigstuhl 12, 69117 Heidelberg}
\affil[d]{INAF - Osservatorio Astronomico di Brera, via E. Bianchi 46, 23807 Merate (LC), Italy}
\affil[e]{SUPA, Institute of Photonics and Quantum Sciences, Heriot-Watt University, Edinburgh EH14 4AS, UK}
\affil[f]{Department of Physics, University of Bath, Claverton Down, Bath BA2 7AY, UK}
\affil[g]{Institute of Microstructure Technology~(IMT), Karlsruhe Institute of Technology~(KIT), Hermann-von-Helmholtz-Platz~1, 76344~Eggenstein-Leopoldshafen, Germany}
\affil[h]{Institute of Photonics and Quantum Electronics~(IPQ), Karlsruhe Institute of Technology~(KIT), Engesserstr.~5, 76131~Karlsruhe, Germany}
\affil[i]{Universit\'e Paris-Saclay, Institut d'Optique Graduate School, CNRS, Laboratoire Charles Fabry, Palaiseau, France}
\affil[j]{Department of Physics, Durham University, South Road, Durham, DH1 3LE, UK}
\affil[k]{LESIA, Observatoire de Paris, Universit\'e PSL, CNRS, Sorbonne Universit\'e, Universit\'e de Paris, 5 place Jules Janssen, 92195 Meudon, France}
\affil[l]{Max-Planck-Institute for Astronomy, K\"onigstuhl~17, 69117~Heidelberg, Germany}
\affil[m]{ITEAM Research Institute, Universitat Polit\`ecnica de Val\`encia, Valencia, 46022, Spain}
\affil[n]{MQ Photonics Research Centre, Department of Physics and Astronomy, Macquarie University, NSW 2109, Australia}

\cftpagenumbersoff{figure}
\cftpagenumbersoff{table} 
\begin{document} 
\maketitle

\begin{abstract}
Direct imaging instruments have the spatial resolution to resolve exoplanets from their host star. This enables direct characterization of the exoplanets atmosphere, but most direct imaging instruments do not have spectrographs with high enough resolving power for detailed atmospheric characterization. We investigate the use of a single-mode diffraction-limited integral-field unit that is compact and easy to integrate into current and future direct imaging instruments for exoplanet characterization. This achieved by making use of recent progress in photonic manufacturing to create a single-mode fiber-fed image reformatter. The fiber-link is created with 3D printed lenses on top of a single-mode multi-core fiber that feeds an ultrafast laser inscribed photonic chip that reformats the fiber into a pseudo-slit. We then couple it to a first-order spectrograph with a triple stacked volume phase holographic grating for a high efficiency over a large bandwidth. The prototype system has had a successful first-light observing run at the 4.2 meter William Herschel Telescope. The measured on-sky resolving power is between 2500 and 3000, depending on the wavelength. With our observations we show that single-mode integral-field spectroscopy is a viable option for current and future exoplanet imaging instruments.
\end{abstract}


{\noindent \footnotesize\textbf{*}NASA Hubble Fellow S. Y. Haffert,  \linkable{shaffert@email.arizona.edu}}

\begin{spacing}{2}   

\section{Introduction}
\acresetall

Exoplanet characterization often makes use of spatially unresolved spectroscopy of transiting planets. During the transit, light from the host star passes through the planet's atmosphere and leaves imprints in the starlight, which can then be analyzed to characterize the planet. This has lead to remarkable characterization of a wide variety of exoplanets \cite{snellen2010orbit, sing2016hotjupiters, kreidberg2019rocky}. The same technique has been applied to measure the direct emission from the planet, either thermal emission or optical reflected starlight, itself instead of the indirect signatures in the starlight \cite{charbonneau2005ecplisetransit, brogi2012taubootis,hoeijmakers2018kelt9b}. These are very challenging observations, but they are within the current limits of technology for hot Jupiters. The main limiting factor for the detection of fainter exoplanets in unresolved spectroscopy is the overwhelming amount of starlight contaminating the planet signal, which has to be removed in post-processing by complex time-series filtering algorithms.

Direct imaging instruments, such as SPHERE \cite{beuzit2019sphere}, GPI \cite{macintosh2014gpi} or SCExAO \cite{jovanovic2015scexao}, are built to spatially resolve planets on large enough orbit from their host stars and often employ \ac{ExAO} systems and advanced coronagraphs. By employing \ac{ExAO} the influence of the star is greatly reduced by spatially resolving the planet, allowing for easier characterization of the planet. The coronagraph is then used in conjunction with an \ac{ExAO} system to suppress the starlight, to enhance the contrast between the star and planet even further. After employing all these optical techniques, there is still residual starlight that leaks through the system due to imperfect \ac{AO} correction and residual phase aberrations that creates speckles which can appear as planets. Image processing techniques are necessary to remove these residual speckles to recover the planet signal.

While several giant planets have been directly imaged and spectroscopically characterized \cite{lagrange2009betapic,bonnefoy2013betapicspectrum,keppler2018discovery,mueller2018characterization} the number of directly imaged planets have been limited so far. The current generation of high-contrast imagers are sensitive to the small population of young self-luminous exoplanets on wide orbits \cite{bowler2015gpoccurence,wagner2019wideoccurence}. More planets could be directly imaged if the sensitivity close to the star is improved. Most of the current direct imaging instruments are limited at angular separations smaller than $10 \lambda / D$ ( 0.4" at the VLT in H-band) \cite{vigan2020shine} by quasi-static-speckles, which are slowly evolving speckles. Post-processing algorithms that depend on spatial diversity, such as \ac{ADI} \cite{marois2006adi}, are not able to remove the \ac{QSS} due to the limited spatial diversity at small angles. 

The combination of high-contrast imaging with high-resolution spectroscopy can resolve the issues that both techniques face. The effective resolving power on which the quasi-static speckles change is mainly dominated by the scaling of the PSF with wavelength. This occurs on a resolving power of $R_{QSS}=\Delta\theta/(\lambda/D)$, with $\Delta \theta$ the angular separation, $\lambda$ the central wavelength of observation and $D$ the telescope diameter \cite{antichi2009bigre}. Observing speckles at 1\,arcsecond with an 8-meter telescope requires a resolving power $R_{QSS}\approx40$. High-resolution spectroscopy can remove the \ac{QSS} efficiently if the spectral features of interest are narrower than the width corresponding to the effective resolution of the \ac{QSS}. This usually holds for atomic and molecular spectral lines, which have an intrinsic resolving power between 100,000 and 200,000. Because these features are so narrow, most spectrographs dilute them to the spectrographs intrinsic resolving power. This sets the requirement that the spectrographs resolving power needs to much larger than the effective resolving power of the speckles $\left(R \gg R_{QSS}\right)$, which happens roughly around $R\approx1000$ and anything above this resolving power is what we consider high-resolution spectroscopy in this work. The combination of high-contrast imaging and high-resolution spectroscopy has been proposed several times already \cite{sparks2002ifu,riaud2007drive, kawahara2014hcs,snellen2015hcihrs,wang2017hdc}, and only recently projects started to add  this capability to high-contrast imaging instruments \cite{rains2018rhea,haffert2018lexi,mawet2018kpic}.

While the high-contrast imaging instruments are lacking high-spectral resolution capabilities, there are several \ac{AO}-fed medium to high-resolution spectrographs (MUSE, SINFONI, OSIRIS, CRIRES). These instruments have been successfully used in the past years to characterize exoplanet atmospheres. The infrared observations of $\beta$ Pictoris b \cite{snellen2014betapic,hoeijmakers2018mapping} and HR8799 \cite{konopacky2013hr8799,barman2015hr8799,dlr2018hr8799,ruffio2019rv} were used to detect the presence of several molecules, including water. Due to the high resolving power of CRIRES \cite{snellen2014betapic} were even able to detect a rotational broadening of $\beta$ Pictoris b. In the visible part of the spectrum MUSE was used to characterize the H$\alpha$ emission of the proto-planet PDS70 b \cite{keppler2018discovery,mueller2018characterization,wagner2018pds70}, and due to the high-sensitivity a second planet was found in the system \cite{haffert2019pds70}. The success of these observations show that it will be worthwhile to add higher resolution integral-field spectroscopy to the current and next generation of high-contrast imagers.

In this work we take advantage of several astrophotonics technologies developed in recent years \cite{leon2012photonic, harris2015photonic, Dietrich2017} to create a large core count single-mode photonic reformatter. With additive manufacturing \cite{dietrich2018printing} a \ac{MLA} is 3D printed on top of the fiber face \cite{Dietrich2017} to efficiently feed the individual cores. An integrated photonic chip is used to rearrange the two-dimensional geometry of the multi-core fiber output into a pseudo-slit that can be dispersed \cite{Thomson:12}. Because the fiber is single-moded, the spectrograph back end can be kept small. The proposed fiber link  can be easily used to add higher resolution spectroscopic capabilities to current generation \ac{ExAO} systems as an upgrade \cite{boccaletti2020sphereplus}.

\begin{figure}
	\begin{center}
        \includegraphics{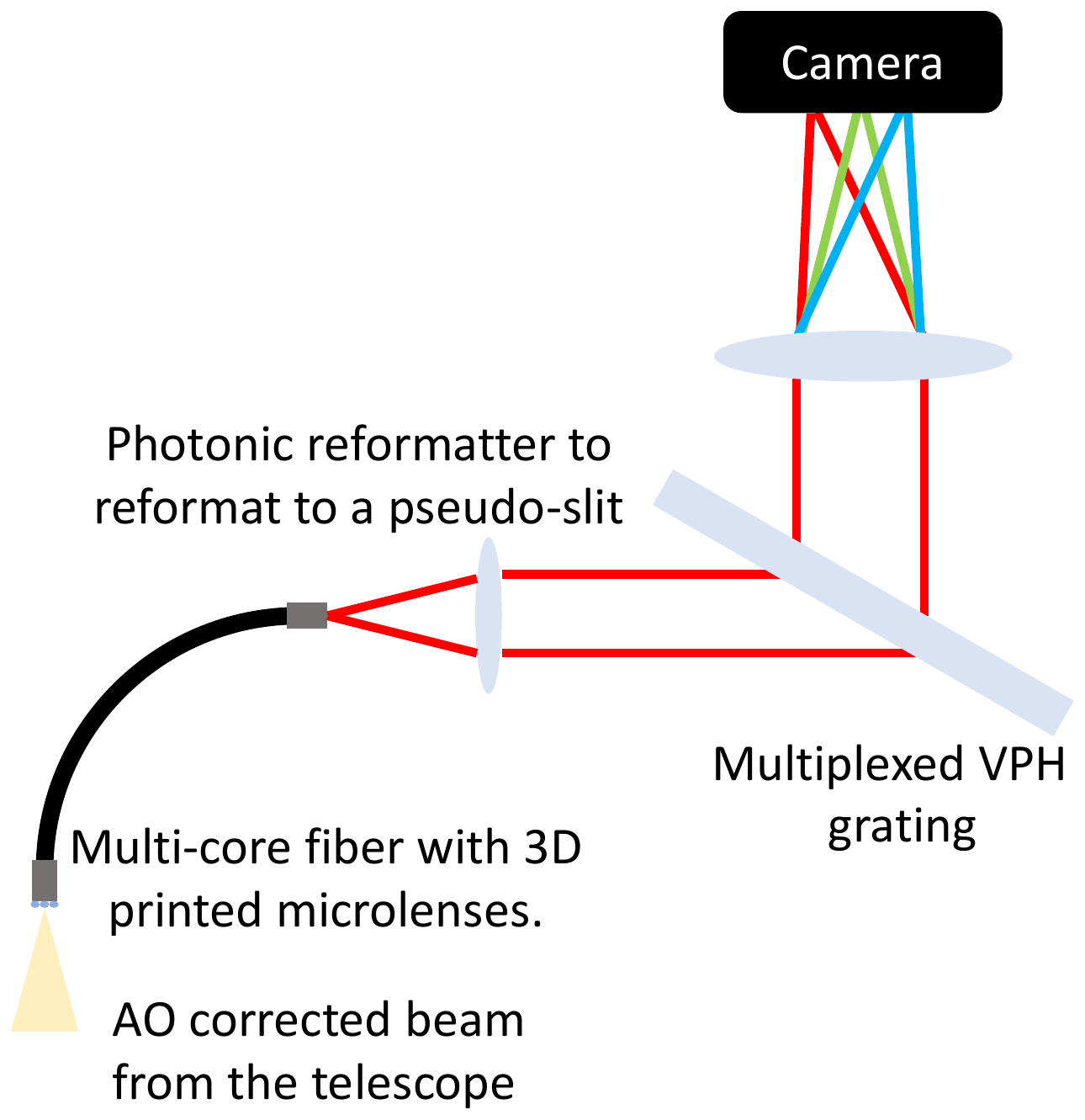}
	\end{center}
	\caption{An overview of the multi-core integral field unit (MCIFU). The telescope beam is imaged onto a microlens array that is written on top of a multi-core fiber. The output of the multi-core fiber is rearanged into a pseudo-slit by the photonic reformatter to make it dispersable. And finally a triple multiplexed grating is used to disperse a broad wavelength range into three orders at a resolving power higher than $R=5000$. The spectrograph itself is used in a first-order manner, where a lens is used to collimate the light onto a grating and a second lens is used to image the spectrograph focal plane. }
	\label{fig:overview}
\end{figure}

The spectrograph in this work uses a \ac{MCF} with 73 cores and has a bandwidth ranging from 1--1.6 $\upmu$m which was set by the properties of the available fiber (lower wavelength limit) and the available detector (upper wavelength limit). This spectral range contains interesting spectral features from molecules such as methane, carbon-monoxide and water and accretion-driven emission lines from hydrogen and helium. The dispersing element of the \ac{MCIFU} is a custom triple stacked \ac{VPHG} that disperses the light into three orders with higher efficiency compared to conventional transmission gratings \cite{zanutta2017spectral}. A schematic of the spectrograph can be seen in Figure \ref{fig:overview}. The prototype \ac{MCIFU} was designed and built during the first half of 2019 and had its first light at the 4.2\,meter \ac{WHT} on La Palma behind the CANARY \ac{AO} system \cite{gendron2016final} in July 2019.

In Section 2 we give a short overview of different image reformatting methods and discuss their advantages and disadvantages for high-contrast imaging. In \ref{sec:fiber_link} we describe the design, manufacturing and characterisation of the novel fiber link, which includes the 3D nano printed \ac{MLA}, the integrated photonic chip and the fiber protection packaging. Section\,4 describes the design and manufacturing of the custom \ac{VPHG}, followed by the opto-mechanical design and characterisation of the spectrograph with the \ac{VPHG} in Section\,5. Section 6 shows the first light results that were achieved with CANARY at the \ac{WHT}. In Section 7 we discuss how to improve the current instrument for future use, after which the paper is summarized and concluded in Section 8.


\section{Comparison of image reformatter concepts for high-contrast imaging}
An image reformatter is an optical component that rearranges an input focal plane into something that can be dispersed by a spectograph without losing spatial information. There are several options to achieve this; image slicers, micro-lenses or fiber-fed spectrographs.
\subsection{Image slicers}
An image slicer uses several reflective elements to reformat the field into a series of mini-slits. The image slicer creates the most efficient packing onto the detector, allowing for the most information to be recorded out of all techniques. But image slicers suffer from spatial-spectral cross-talk because adjacent spatial pixels along each slit are not separated on the detector. Another down-side is that the shape of the \ac{LSF} changes depending on the the mini-slit illumination, which can occur for example due a slight tip or tilt of the input beam. \cite{bacon1995tiger} If \ac{HRS} is needed to gain a large amount of contrast in post-processing to characterize planets, it is necessary to have a stable \ac{LSF} because the \ac{HRS} technique effectively searches for \ac{LSF} variations across the field. If the contrast ratio that has to be bridged with respect to the local stellar halo is not very large ($10^1-10^3$), image slicers may be a good choice as previous observations have shown \cite{thatte2007hciifu, haffert2019pds70}.

\subsection{Micro-lens arrays}
\ac{MLA} based \acp{IFU} use a micro-lens array in the focal plane to sample the field. The \ac{MLA} based \ac{IFU} is the current choice for spectroscopy on  \ac{ExAO} systems \cite{groff2015charis,skemer2015ales,beuzit2019sphere} and is even the standard observing mode for GPI \cite{macintosh2014gpi}. A \ac{MLA} \ac{IFU} is easy to implement and is efficient when few spectral samples are required for large fields \cite{antichi2009bigre}. Another added benefit is that there are several ways to reduce cross-talk between spatial-pixels \textcolor{black}{by adding pinhole masks behind the \ac{MLA} or aperture masks inside the spectrograph} \cite{antichi2009bigre}. The reduced cross-talk allows the \ac{MLA} based \acp{IFU} to reach very deep contrast ratios \textcolor{black}{of $10^4-10^6$} \cite{vigan2015sirius, beuzit2019sphere}. The drawback of the \ac{MLA} \acp{IFU} is that it is not possible to measure many spectral bins per spatial pixel (spaxel). Due to diffraction of the micro-lenses the spot size is $\lambda F_{\mathrm{mla}}$ with $\lambda$ the wavelength and $F_{\mathrm{mla}}$ the focal ratio of the micro-lens. \textcolor{black}{Each spot requires an area of $\lambda^2 F_{\mathrm{mla}}^2$, while the total available area for each spaxel is the area of a single micro-lens, $D_{\mathrm{mla}}^2$.} \remove{And for each spaxel there is an area available of $D_{\mathrm{mla}}^2$.} That means that the maximum number of spectral bins is, 
\begin{equation}
    N_{\lambda} = \frac{D_{\mathrm{mla}}^2}{\lambda^2 F_{\mathrm{mla}}^2}.
\end{equation}
The \textcolor{black}{relative} bandwidth of the spectrograph is equal to the number of spaxels divided by the resolving power, $R=\lambda/\delta\lambda$, of the spectrograph, 
\begin{equation}
    \frac{\Delta \lambda}{\lambda } = \frac{N_{\lambda}}{\eta R} = \frac{D_{\mathrm{mla}}^2}{\eta R \lambda^2 F_{\mathrm{mla}}^2}.
\end{equation}
Here $\Delta \lambda$ is the spectrograph bandwidth and $\eta$ is the detector filling efficiency. The usual range of $\eta$ is between $1/2$ and $2/3$ to ensure either a separation of 1 or 2 resolving elements between the spectra, respectively.

The typical micro-lens diameter is on the order of 300\,$\upmu$m and for the \ac{NIR} a central wavelength of $1.4 \upmu$m is reasonable. If the micro-lenses have a focal ratio of 5 the total bandwidth that can be observed at a resolving power of 10000 with a single exposure is 16\%. The bandwidth reduces to 8\% if the detector separation is included. For high-resolution spectroscopy ($R=100000$) the bandwidth becomes a factor 10 even smaller. The single-shot bandwidth is a crucial parameter for the \ac{HRS} technique because the \ac{SNR} of the template matching is proportional to the bandwidth and the integration time, $\mathrm{SNR} \propto \sqrt{\Delta \lambda \Delta t}$ \cite{sparks2002ifu,snellen2015hcihrs}. If the single-shot bandwidth is not high enough \textcolor{black}{conventional imaging} techniques, \textcolor{black}{such as ADI}, may be more efficient for exoplanet imaging \cite{males2018predcon}. The MLA approach would be very suitable to target a small bandwidth around emission lines, such as H$\mathrm{\alpha}$, at high-resolution to search for proto-planets.

\subsection{Fiber-based IFUs}
A field can be reformatted by feeding each spaxel into a fiber. A fiber bundle has the most flexibility of all reformatting techniques, the input and output can be rearranged completely independent from each other. The fiber-based \ac{IFU} was therefore also the first \ac{IFU} concept to be used on-sky \cite{vanderriest1980fifu}. Most fiber-based \acp{IFU} until now have used \ac{MMF}. \textcolor{black}{These fibers can capture all light due to their large core, but have a low fill fraction due to their large cladding diameters. This problem has been reduced by feeding the fibers with a micro-lens array to increase the field \cite{Allington-Smith:1997}}. The past few years has shown that the mode filtering capabilities of \acp{SMF} can be used to create coronagraphs with smaller inner-working angles or higher throughput \cite{mawet2017hdc, por2018SCARI, ruane2018vfn, coker2019mos}.

Additionally \acp{SMF} can reject random speckles from the \ac{ExAO} system to increase constrast \cite{mawet2017hdc}, which has been used before in the interferometry community to make it easier to interfere different beams \cite{petrov2007amber,gravity2017}. Because single-mode fibers only propagate a single-mode, the LSF is also very stable which will make it easier to calibrate the spectra. Single-mode spectroscopy is currently pursued in the radial velocity field \cite{crepp2016ilocator}, because the \ac{LSF} from multi-mode fibers is not stable enough \cite{baudrand2001modalnoise}. And as a final benefit because \acp{SMF} only allow the propagation of a single-mode, any incoherent source that consists of many modes will be attenuated \cite{fohrmann2015smfthermal}. This property is very desirable especially in the infra-red where the thermal background from the sky is high and limits the \ac{SNR} of most observations. Therefore, a \ac{SMF} bundle provides a significant amount of benefits over other manners of image slicing making it a very attractive option for high-resolution broad-band integral-field spectroscopy. It is difficult to quantify the gain of an \ac{SMF} \ac{IFU} over other methods without complete end-to-end simulations, which include complete modelling of the instrument, calibration and observing modes. \textcolor{black}{A downside of \acp{SMF} is that the mode filtering capability is restricted to focal plane samplings of 1 $\lambda / D$ per spaxel or larger. If the spatial sample density is increased the mode filtering becomes less effective, which removes the main benefit of the \ac{SMF}. From this it follows that an optimally designed \ac{SM} \ac{IFU} will most likely not Nyquist sample the focal plane, which will remove the ability to use post-processing techniques like \ac{ADI}. But with the addition of high-resolution spectroscopy it is possible to apply post-processing techniques like Molecule Mapping \cite{hoeijmakers2018mapping} or High-Resolution Spectral Differential Imaging (HRSDI) \cite{haffert2019pds70} to discover and characterize exoplanets.}

\subsection{The MCIFU reformatter}
To reach a high field filling fraction the \acp{SMF} have to be fed with micro-lenses \cite{corbett2009smf, por2018SCARI}. \acp{SMF} have very \textcolor{black}{strict} requirements on the alignment of the fibers for efficient injection\cite{jovanovic2017injectsmf}, even more if they need to be combined with coronagraphy \cite{haffert2018SCARII}. Currently the only \acp{SMF}-fed \ac{IFU} is RHEA \cite{rains2016rhea}, which uses a bundle of \acp{SMF} that are fed by a bulk micro-lens array. Accurate alignment of the \ac{SMF} bundle behind the \ac{MLA} is very difficult \textcolor{black}{with typical depths of focus of 10-20 $\upmu$m and lateral alignment tolerances $\le1\upmu$m} \cite{rains2018rhea}. And even in case that the individual fibers would be aligned for high injection efficiency, getting the output of the fibers aligned in the same plane of focus for the spectrograph is still difficult\cite{rains2018rhea}. Both issues can be solved by using multi-core fibers, which contain several independent \ac{SM} cores inside a single fiber. This ensures that the fibers are in the same input and output plane. But because of the small size of the \acp{MCF} it is difficult to feed them with a bulk micro-lens array \textcolor{black}{and reach the required alignment accuracy}. We will use in-situ 3D printing of the micro-lens array directly on top of the fiber face \textcolor{black}{, which will allow for much more freedom.} \textcolor{black}{However,} due to fixed core spacing it is difficult to disperse the output over a broad bandwidth if there are many cores\cite{betters2014pimms}. \textcolor{black}{The pitch constraint} is similar to the constraint of micro-lens based \acp{IFU}s \cite{bacon1995tiger, antichi2009bigre}.

\section{The single-mode multi-core fiber link}
\label{sec:fiber_link}

The fibre-link allows the \ac{PSF} from the \ac{AO} system to be coupled to the spectrograph. A description of all the different elements of the \ac{MCIFU} is presented in Figure \ref{fig:HWfig1}. Figures \ref{fig:HWfig1}(a) and (b) show a schematic and photograph of the full \ac{MCIFU}, with the five main components; the \ac{MLA} \,(c), the \ac{MCF} (d), the MCF glued inside a custom V-groove (e), the reformatter (f-g), and  an ultrafast laser fabricated mask to block unguided light in the reformatter (h). Finally, Figure \ref{fig:HWfig1}(f,g) represents a schematic representation of the reformatter with the colour scheme to distinguish each row of waveguides.

The fiber link was assembled by first manufacturing and connectorizing the \ac{MCF} and chip, then aligning and gluing to the reformatter and mask which were manufactured using \ac{ULI}. This was then packaged using off-the-shelf and 3D printed components and finally the microlenses were printed on the \ac{MCF} which was secured in an FC-PC connector. The following subsections are arranged in the order of manufacture.

\subsection{Fiber}
The \ac{MCF} is comprised of 73 step-index Ge-doped cores manufactured using the common stack-and-draw fibre fabrication technique (Figure \ref{fig:HWfig1}(d)). The \ac{MCF} has a outer cladding diameter of 560\,$\upmu$m. Each individual core has a \ac{NA} of 0.14 and a diameter of approximately 5.3 $\upmu$m, \textcolor{black}{which results in a \ac{MFD} of approximately 8.2 $\upmu$m at a wavelength of 1.4 $\upmu$m}.  The \ac{SM} cutoff of each core is approximately 970 nm. The cores are spaced by 41 $\upmu$m, this means they are separated by approximately \textcolor{black}{5} times the \ac{MFD}, ensuring negligible cross coupling between the cores.

\subsection{Reformatter and mask}

\begin{figure}
	\begin{center}
        \includegraphics{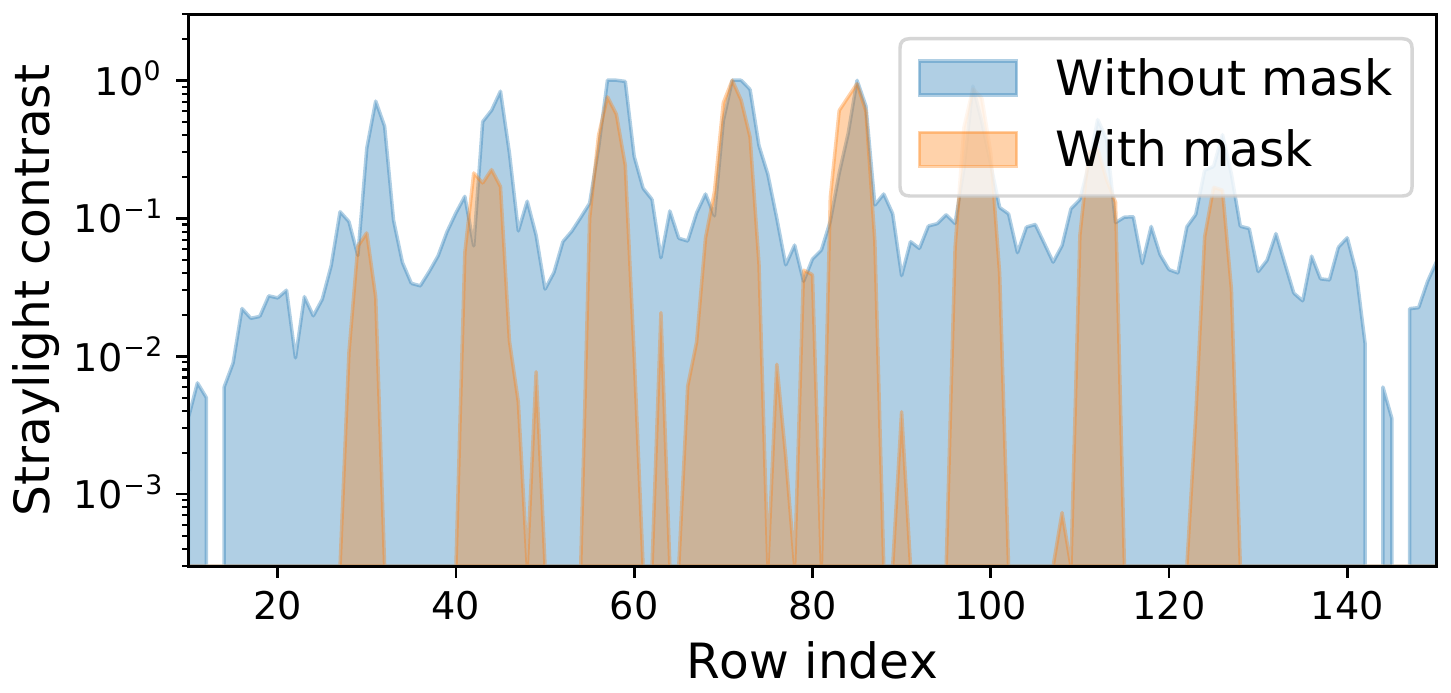}
	\end{center}
	\caption{A slice along one of the rows of the reformatter showing the scattered light with and without the chromium mask. Here, the input microlens array is overfilled, specifically to simulate background light in astronomical observations. The blue shaded area, shows where the scattered light is still detectable between the waveguides, the red shaded area shows the stray light with the mask. The row index is the pixel index along the extracted slice.}
	\label{fig:HWcomparison}
\end{figure}

At the spectrograph end, the \ac{MCF} is directly butt-coupled to a \ac{ULI} fabricated three-dimensional waveguide reformatter, which spatially reformats the \ac{MCF} cores into a suitable arrangement, such that the dispersed spectra from each waveguide do not overlap with each other on the detector. To minimise undesirable effects due to scattered / unguided light not contained within the reformatter cores, a mask was fabricated using \ac{FLICE}. This mask consisted of a fused silica fixture with holes precisely positioned to align with the output waveguides. Measurements taken after coating the silica with a layer of chromium, show that the chromium mask is blocking the stray light by a factor of 10 to a 100 (see Figure \ref{fig:HWcomparison}). 
\begin{figure}
	\begin{center}
        \includegraphics[width=\textwidth]{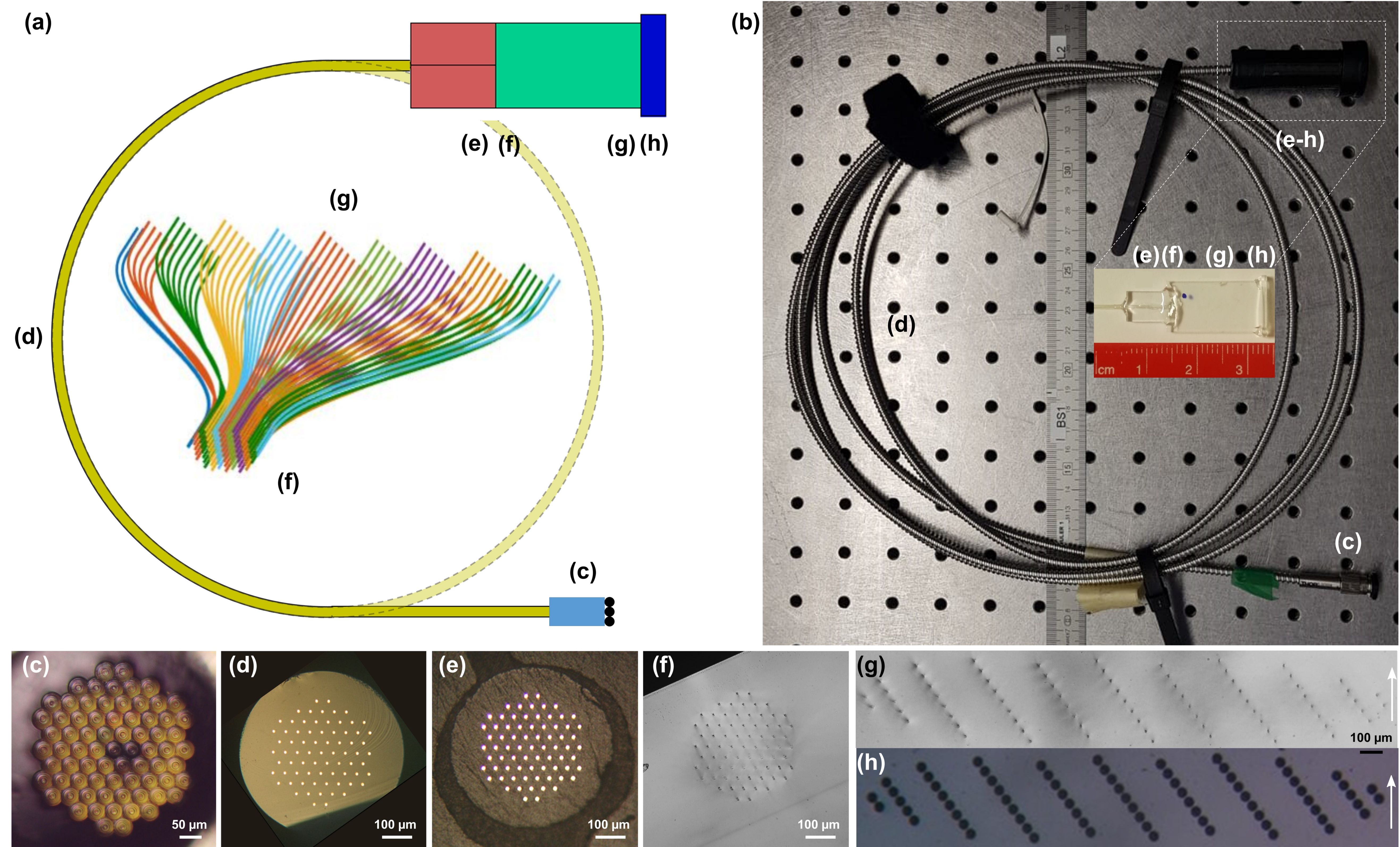}
	\end{center}
	\caption{The MCIFU fiber link, with important sections shown. (a) A schematic representation of the fiber link, (b) photograph of the complete packaged fibre-link comprising all the elements from the microlens array, (c) the 3D printed microlenses, (d) the bare fiber, (e) the bare fiber in a v-groove, (f) the input of the chip reformatter. (g) The output of the reformatter (h) The output mask. The white arrow in (g/h) indicates the dispersion direction. Inset in (b): scale picture of the full reformatter.}
	\label{fig:HWfig1}
\end{figure}

To separate the spectra from each \ac{MCF} core and avoid any overlap \textcolor{black}{after dispersion}, the quasi-hexagonal \ac{MCF} shape must be reformatted (Figure \ref{fig:HWfig1}(g)). \textcolor{black}{Existing reformatters have generally chosen a linear pseudo-slit output pattern \cite{Thomson:12,Spaleniak:13,harris2015photonic}, however for the \ac{MCIFU} we chose a staggered slit. This has the advantage of increasing the spacing between adjacent cores, reducing cross coupling, whilst also reducing the translation of the waveguides within the glass. The reformatted pattern was created to give a spacing of 30 $\upmu$m between the individual spectral traces of each core on the spectrograph.} To do so, \ac{ULI} was used to inscribe a reformatter in a 20 $\times$ 10 $\times$ 1 mm borosilicate glass substrate (Eagle XG) \cite{maclachlan2016efficient}. The inscription laser source is a MenloSystems BlueCut fibre laser emitting at 1030 nm a 500-kHz train of 350 fs pulses which are focused within the substrate with a 0.55 \ac{NA} aspheric lens. Each waveguide was inscribed using a substrate translation speed of 8 mm/s and 19 scans of the laser focus with an inter-scan separation of 0.2 $\upmu$m. This resulted in highly symmetric \ac{SM} waveguides with an \ac{NA} of $\approx$0.11 and \ac{MFD} of $\approx$7.3\,$\upmu$m at the 1/$\mathrm{e^{2}}$ beam size of a Gaussian beam at 1310 nm. \ac{ULI} has a limit in the refractive index contrast that can be achieved, which resulted in a lower \ac{NA} of the \ac{ULI} waveguides as compared to the \ac{NA} of the \ac{MCF}. This will lead to coupling losses at the interface of the reformatter and the \ac{MCF} \textcolor{black}{that represents a significant part of the global insertion losses of the reformatter}. The pulse energy providing the highest throughput waveguides at the initial depth of 470 $\upmu$m was 128 nJ and adjusted to 105 nJ for depths lower than 130 $\upmu$m to compensate for spherical aberrations on the beam.

The waveguides were characterised by two different laser sources: a 1310 nm laser and a supercontinuum with an 1100\,nm bandpass filter (with a \ac{FWHM} of 10\,nm). The two different sources were coupled into an SMF-28 optical fibre and characterization performed by butt-coupling the SMF-28 output to a single core of the \ac{MCF}. The \ac{MCF} was then butt-coupled to each individual waveguide of the reformatter with the aid of an index matching fluid. By this way, we ensure to work with the exact mode field diameter to characterize the waveguides. \ac{SM} behaviour of the waveguides in the J-band, between 1.1 and 1.3\,$\upmu$m was observed. The throughput was characterized with the ultra-stable laser at 1310\,nm. With this optimized set of parameters, the throughput of the straight waveguides can reach 67\,\%, \textcolor{black}{corresponding to a global insertion losses of 1.7 dB}. To create the reformatter, an investigation was performed to optimise bend radius and minimise transmission losses. The bending losses are negligible for a bend radius in the propagation direction of $>$15\,mm, deviating by a distance of 1 mm (corresponding to the extreme edge waveguides), and we have chosen a bend radius of 18 mm.

The final reformatter was permanently glued to the \ac{MCF} using a UV curing adhesive. To provide a larger gluing surface to bond the fibre to the reformatter a fibre support chip was fabricated with a custom hollow-cylindrical V-groove to house the \ac{MCF}. This was created by the process of \ac{FLICE}, in an (8 $\times$ 6 $\times$ 2 mm) fused silica substrate \cite{ross2018}. The V-groove shape was defined by the process of \ac{ULI} with 200 nJ pulses from the Bluecut fibre laser at a repetition rate of 250 kHz and a subsequent wet chemical etch in 8 mol.L$^{-1}$ potassium hydroxide solution at 85 degrees to remove the excess material.

Finally, a mask was developed to block stray and scattered light at the reformatter output while allowing the light from the individual waveguides to propagate as desired. The scattered light in the \ac{ULI} glass substrate can present a significant challenge for astronomical instruments trying to overcome the high contrast between the observed star and the planet \cite{jovanovic2012starlight}. The mask (Figure \ref{fig:HWfig1}(h)) was created by the process of \ac{FLICE} in a similar manner to the fibre support chip from a 2 $\times$ 12 $\times$ 2 mm fused silica substrate. The high precision stages enabled a rectangular slot to be created to snugly fit the reformatter within this mask, with 30\,$\upmu$m diameter through holes precisely passively aligned to the waveguide positions (Figure \ref{fig:HWfig1}(g)). A 120\,nm layer of chromium metal was deposited on the outer surface of the mask by electron-beam physical vapour deposition. This process creates an opaque component that removes the scattered light present in the glass between the waveguides. Figure \ref{fig:HWcomparison} illustrates the removal of scattered light between the \ac{ULI} waveguides without and with the opaque mask, respectively in blue and red. \textcolor{black}{The scattered light was measured by taking a slice along one of the rows of the reformatter.}

\begin{figure}
    \begin{center}
        \includegraphics{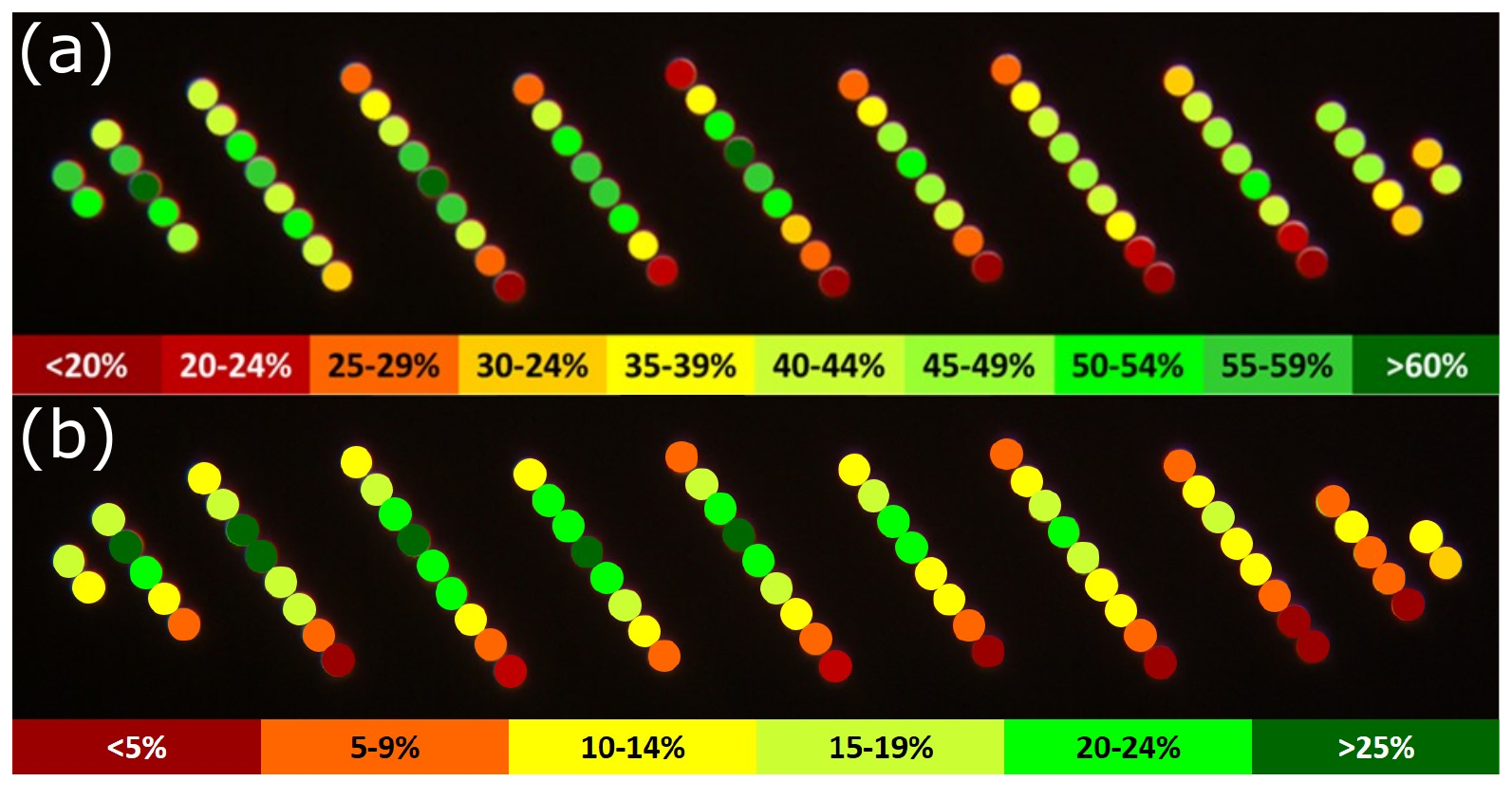}
	\end{center}

\caption[Two numerical solutions]{Illustration of the measured monochromatic (1310\,nm) throughput results for the fiber link. (a) shows the throughput for the 73 individual cores through the fiber, the reformattter and the chromium mask. (b) as for (a) but for the full fiber link, including the microlenses. The colour bars represents the throughput ranges for each figure.}
\label{fig:HWfig2} 
\end{figure}

In Figure \ref{fig:HWfig2}, we present the end-to-end throughputs from the \ac{MCF} to the mask, at 1310 nm, before the \ac{MLA} was added. In Figure \ref{fig:HWfig2}, one can see clearly that the fibre-link has lost a significant part of the throughput on the top/bottom edges of the reformatter. These losses come from a slight variation of the input waveguide positions of the reformatter (Figure \ref{fig:HWfig1}(f)) in comparison of the \ac{MCF}, inducing mode-field diameter mismatching.

The whole output device was then encased with a combination of off-the-shelf components and 3D printed parts, and the \ac{MLA} was secured within an FC-PC connector to create a robust component for use at the telescope (Figure \ref{fig:HWfig1}(b)).

\subsection{Microlens array}

\begin{figure}
	\begin{center}
        \includegraphics{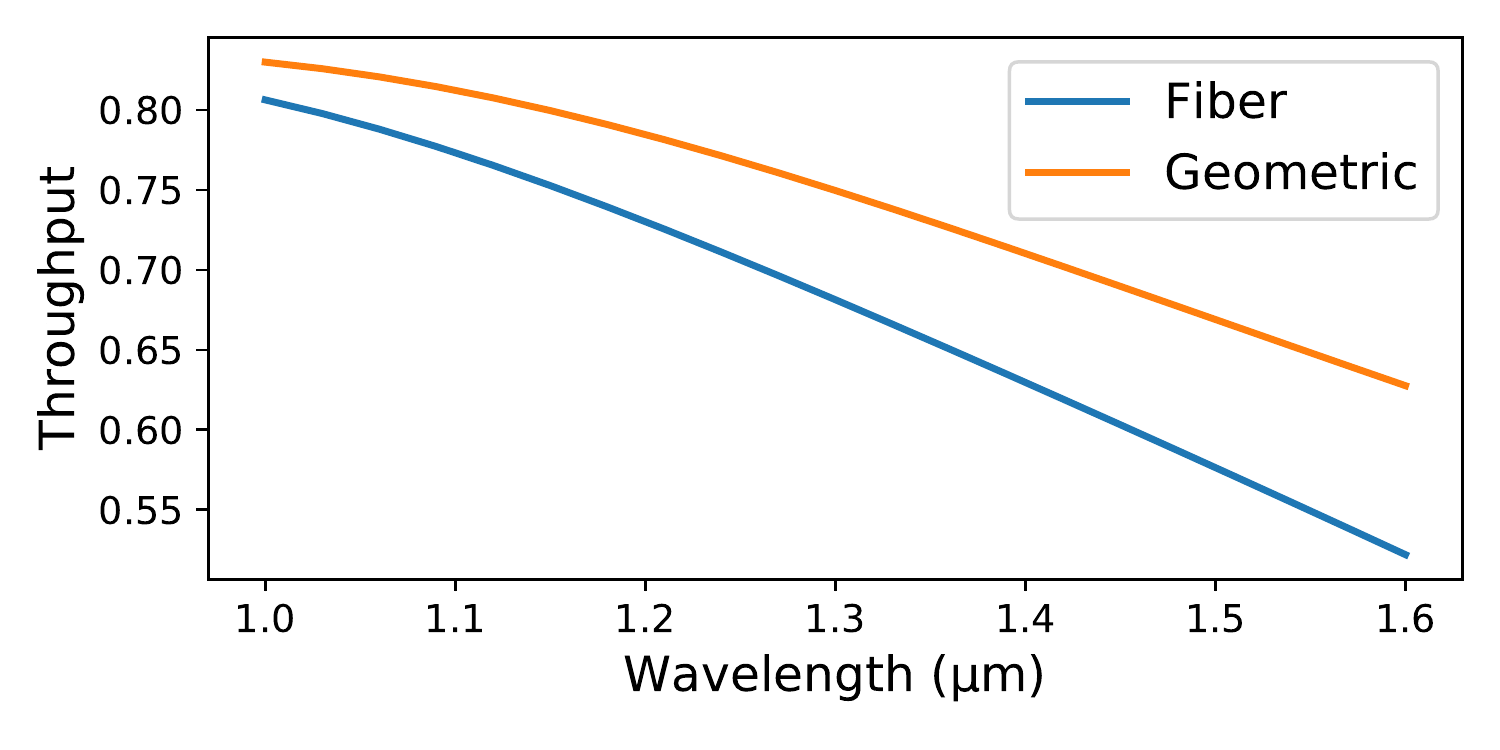}
	\end{center}
	\caption{The amount of on-axis light that couples into a single core of the \ac{MCF} as function of wavelength within the spectrograph bandwidth. The fiber throughput is the amount that couples into a single core, while the microlens throughput is the amount of light that enters the microlens. With the microlens array we can couple between 80\% and 95\% of the light that falls on a single microlens. }

	\label{fig:design_coupling_efficiency}
\end{figure}

The surface shape of the \ac{MLA} was optimized for coupling efficiency using the physical optics propagation module in Zemax. To simulate the Airy pattern of a telescope we used an unobstructed aperture with a diameter of 1 m. The beam was focused by a paraxial lens with a focal length that created a focal ratio of \textcolor{black}{22}, allowing a 1.5\,$\lambda/D$ sampling per microlens of the focal plane at 1.3\,$\upmu$m. The microlens surface was then modelled as a single lens with a hexagonal aperture made from IP-DIP\cite{gissibl2017ipdip}. We tried several freeform shapes to increase the coupling efficiency but found that there was very little improvement, less than a percent point, compared to purely spherical surfaces. Therefore we chose a spherical surface for the final design. We found that a radius of curvature of 74.5 $\upmu$m with a 205 $\upmu$m height provided the highest wavelength averaged coupling efficiency of 67.5\% to a single core of our \ac{MCF}. Figure \ref{fig:design_coupling_efficiency} shows the on-axis coupling efficiency as function of wavelength.The proposed microlens design has a theoretical wavelength dependent throughput between 50\% and 81\%. The limiting factor in the \textcolor{black}{throughput} \remove{coupling} is the fraction of the Airy Pattern that a single microlens captures. Due to scaling of the Airy Pattern with wavelength, longer wavelengths will capture a smaller fraction of the \ac{PSF}. \textcolor{black}{The orange line in Figure \ref{fig:design_coupling_efficiency} shows the fraction of the transmitted flux for a micro-lens as a function of wavelength. It is clear that the coupling to the \ac{SMF} cores is between 80,\% and 95,\% efficient across the spectrum, which shows that there will be little to gain from more complicated designs for on-axis objects. }

The \ac{MLA} was \textit{in-situ} printed on the flat facet of the FC-PC connector in which the cleaved \ac{MCF} was manually glued and then polished. The lenses were printed in a single block and in a single two-photon lithography step out of the commercial negative-tone photoresist IP-Dip \cite{nanoscribe}. The structures were defined by an in-house built lithography machine, equipped with a 780\,nm femtosecond laser. An in-house developed machine software was used for high-precision alignment and writing with high shape fidelity. Illumination of the fiber back end with a red LED together with machine vision was used to detect all 73 cores of the \ac{MCF} and align the individual lenslets to the cores. Therefore, the full 3D-model is generated only after core detection in order to be able to compensate any slight location and pitch variation of the individual cores of the \ac{MCF}. \textcolor{black}{The individual models of the lenslets are then merged and at places of overlap, due to the slight spatial variation in the \ac{MCF} cores, the highest surface is chosen.} Slight tilts of the fiber facet due to mounting are detected as well and the structures corrected accordingly. The writing distances between subsequent lines and layers, \textit{i.e.}, both hatching and slicing distance, were set to 100 nm. \textcolor{black}{No \ac{AR} coating was added to the micro-lenses. This adds some losses due to Fresnel reflection, which are roughly 5\%.}

\subsection{Performance of the fiber link}

With the whole system assembled and packaged the individual cores were tested for throughput. This was again performed at 1310 nm to be consistent with earlier tests. The light sources were fed through a single mode fibre and then to a collimating lens (Thorlabs AC254-200-C). The collimated beam was stopped down by an aperture stop to simulate the same focal ratio of 22 beam that the foreoptics \textcolor{black}{of the experiment with CANARY} will supply to the microlenses. Finally a lens (Thorlabs AC254-100-C) was used to focus the light onto the microlenses. To simulate the on-sky conditions the input light was first optimised for position, focus, tip and tilt using a central lens. This was optimized and checked at several points across the lens array and then used as the reference. From this point only \textcolor{black}{the horizontal ($x$) and vertical ($y$) position} were changed, maintaining a global reference for all the microlenses. This means any deviation from perfect will have reduced the throughput of the individual lenses, however it better matches the expected on sky conditions. \textcolor{black}{This simple setup creates a diffraction-limited beam and shows the wavefront error free throughput.}

The recorded throughputs are shown in Figure \ref{fig:HWfig2}. This shows that the central cores performed best, with significant reduction at the edges. Causes and solutions are discussed in Section \ref{sec:Discussion}. 

\section{The triple stacked Volume Phase Grating}
\begin{figure}
	\begin{center}
        \includegraphics[width=\textwidth]{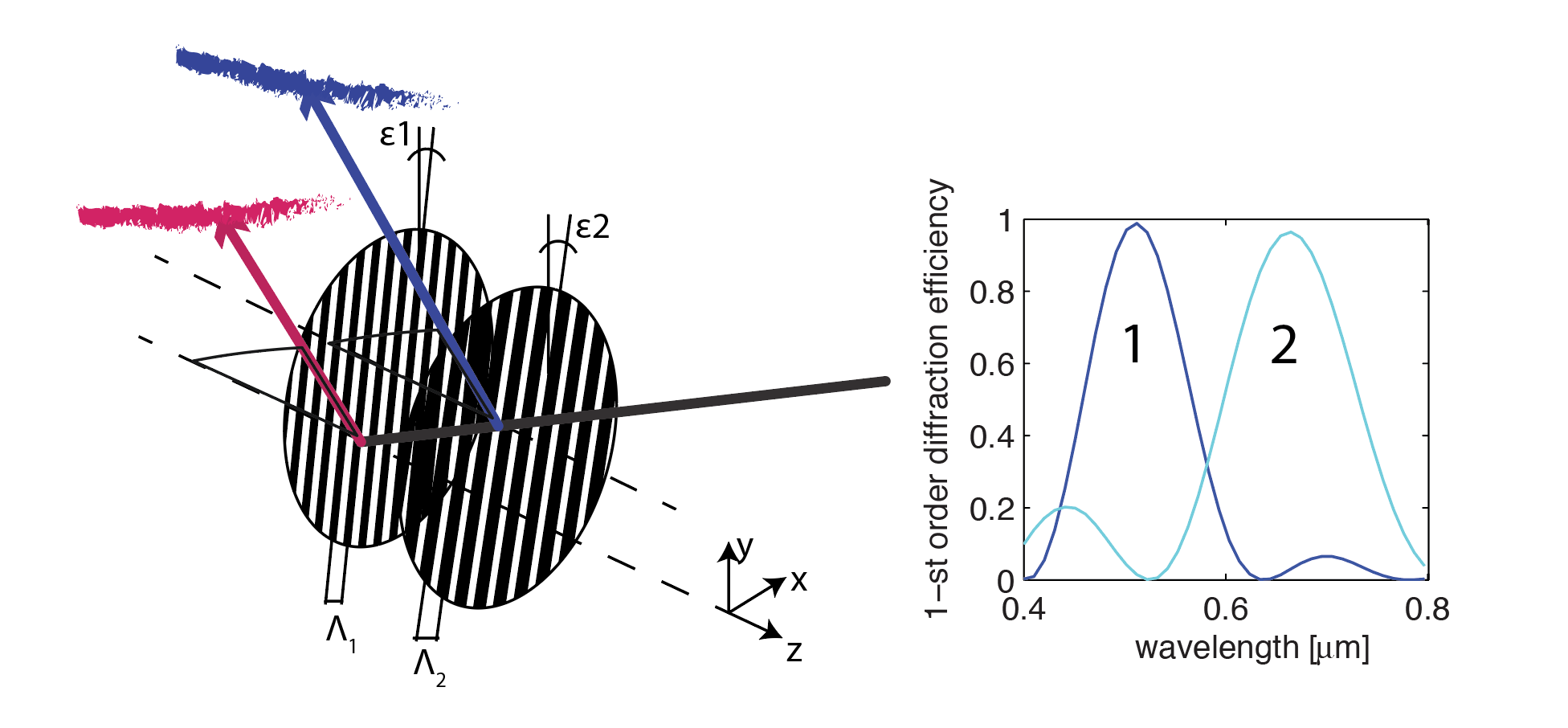}
	\end{center}
	\caption{Scheme of the spectral-multiplexing concept. For simplicity a two-stacked \ac{VPHG} is presented, each creating a separate spectrum which are shown in different colours. The inset shows the simulated diffraction efficiencies of the two layers for an example design.}
	\label{fig:multi1}
\end{figure}

The ultimate goal of the instrument is to have a spectral resolving power between 5,000 and 10,000, which can be readily accessible with transmissive volume phase gratings (used in the 1st diffraction order with the present design). Despite this, even if \acp{VPHG} can be made with very high diffraction efficiency, the disadvantage of a classical, single first-order grating is that the light will be smeared out in a single spectrum, thus wasting usable portions of the detector. Moreover, for securing the large bandwidth at the target resolution, it is required to use a very wide sensor, which is very expensive and not readily available.
A possible solution is to multiplex the \ac{VPHG} in layers as reported in \cite{zanutta2017spectral}. In this way, each grating can diffract a different part of the wavelength range towards the same direction so the average efficiency is improved by a wide margin, by having more than one efficiency maximum. To separate the spectra on the detector plane, the gratings are rotated with respect to each other, with a designed angle set in order to have clearly separated spectra and output signals which are as horizontally dispersed as possible (see Figure \ref{fig:multi1}).

\subsection{Design}
Considering the working wavelength range for this application of 1000 -- 1600 nm, the geometrical conditions have been chosen with the aim to obtain three 1st-order diffraction efficiency curves with maxima at around 1100, 1300 and 1500 nm, ensuring a high efficiency at each wavelength.
The incidence angle defined in the spectrograph design is 21.5$^\circ$. This angle is shared by the three dispersing layers and the grating line density has been chosen to match the Bragg condition at the central wavelength for each of the three gratings.

\begin{figure}
	\begin{center}
        \includegraphics{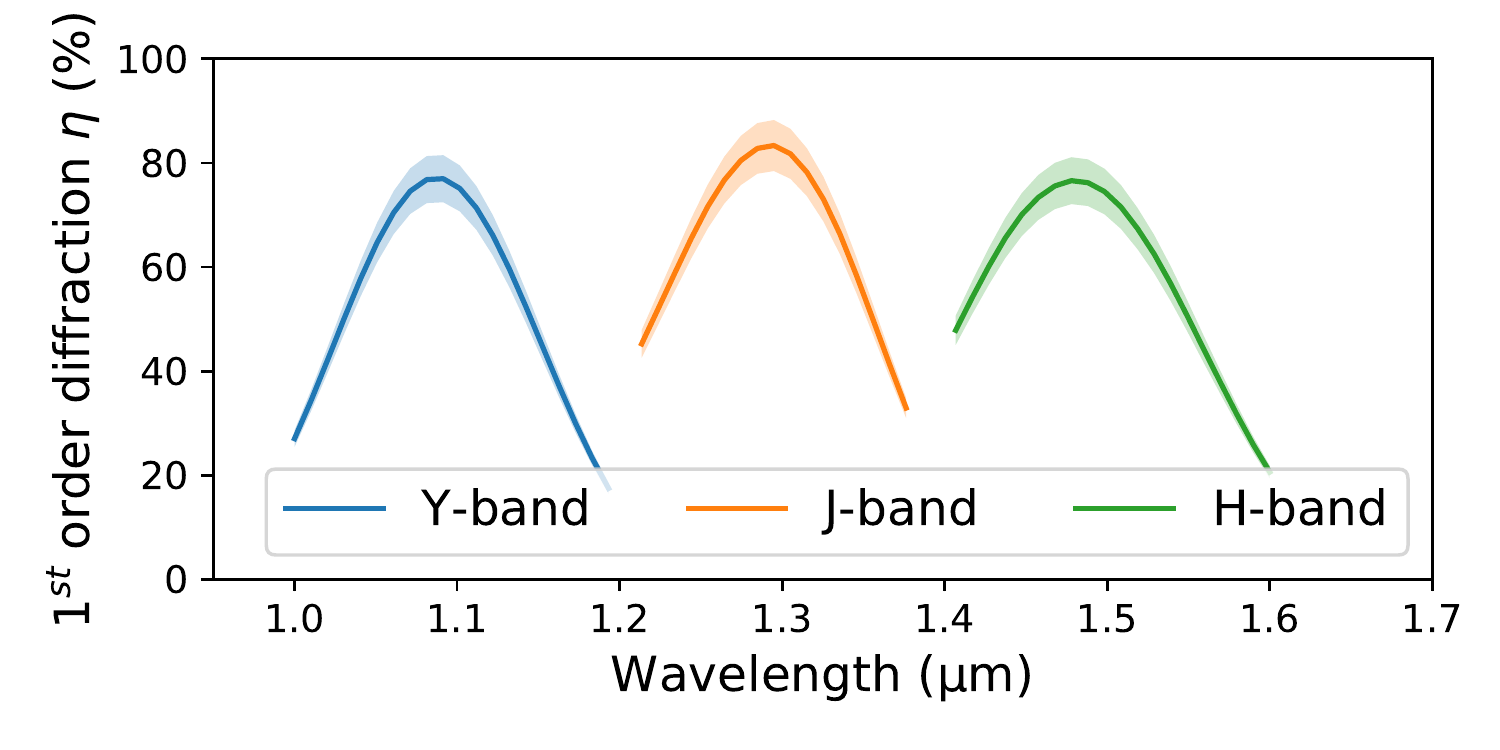}
	\end{center}
	\caption{Simulation of the combined diffraction efficiency of the multiplexed \ac{VPHG} for an incidence angle of 21.5$^\circ$ for unpolarized light. The shaded area present the minimum and maximum diffraction efficiency prediction for the manufacturing process. Material absorption and reflection losses have been estimated and taken into account.}
	\label{fig:vph_simu}
\end{figure}

The efficiency and bandwidth of each single layer (and thus the refractive index modulation strength, $\Delta n$, of the grating and thickness of \acp{VPHG} respectively) has been calculated and optimized by means of a \ac{RCWA} based script. The ad-hoc code takes into account the interaction between the diffractive layers and computes the final throughput of the multiplexed element, see Figure \ref{fig:vph_simu}. The losses due to reflection, as well as those due to materials absorption and scattering have been considered in the simulations in order to have a reliable prediction of the efficiency curves.
In Table \ref{tab:multi} are reported the final design parameters of the stacked optical element.

\begin{table}[ht!]
\caption{Main features selected for the three designed diffraction gratings.}
\centering
 \begin{tabular}{|| m{1cm} | m{2.5cm} m{2cm} m{2cm} m{2cm} m{2cm} ||} 
 \hline
 Layer $\#$ & Central wavelength [nm] & Index modulation $\Delta n$ & Layer thickness [$\upmu$m] & Grating pitch [lines/mm] & $\epsilon$ rotational angle [$^\circ$]\\
 \hline\hline
 1 & 1100 & 0.022 & 28 & 668 & -3 \\ 
 \hline
 2 & 1300 & 0.020 & 34 & 565 & 0 \\
 \hline
 3 & 1500 & 0.010 & 50 & 490 & 3 \\
 \hline
\end{tabular}
 	\label{tab:multi}
\end{table}

\subsection{Manufacturing}

The choice of the materials with different thickness has been based on the $\Delta n$ requirements along with the width of each 1$^{\rm{st}}$ order diffraction and the line density required for each grating. The three gratings have been produced using the Bayfol$^\text{\textregistered}$ HX material. This photopolymeric film has been chosen due to its ability to address precisely the required $\Delta n$ values, simply selecting the proper writing conditions (e.g. laser power density) \cite{zanutta2016photopolymeric}.
Moreover, the solid layers can be used in a configuration that requires only one glass substrate in between the layers, resulting in a thinner diffractive element. Each \ac{VPHG} layer composing the final device has been produced individually by means of the holographic setup based on a 532 nm \ac{DPSS} laser. Every photopolymeric grating is supported by a BK7 round window (1 inch), the front and the back ones possess \ac{AR} coating on the outer face (optimized from 1 -- 1.6 $\upmu$m at 21.5$^\circ$).

After the characterization of each layer, they were coupled with index matching fluid (cedar oil) and aligned according to the tilt angles specified in the design phase.
To hold and block firmly the three layers, an ad-hoc 3D-printed plastic (PLA) shell was used.

\begin{figure}
	\begin{center}
        \includegraphics{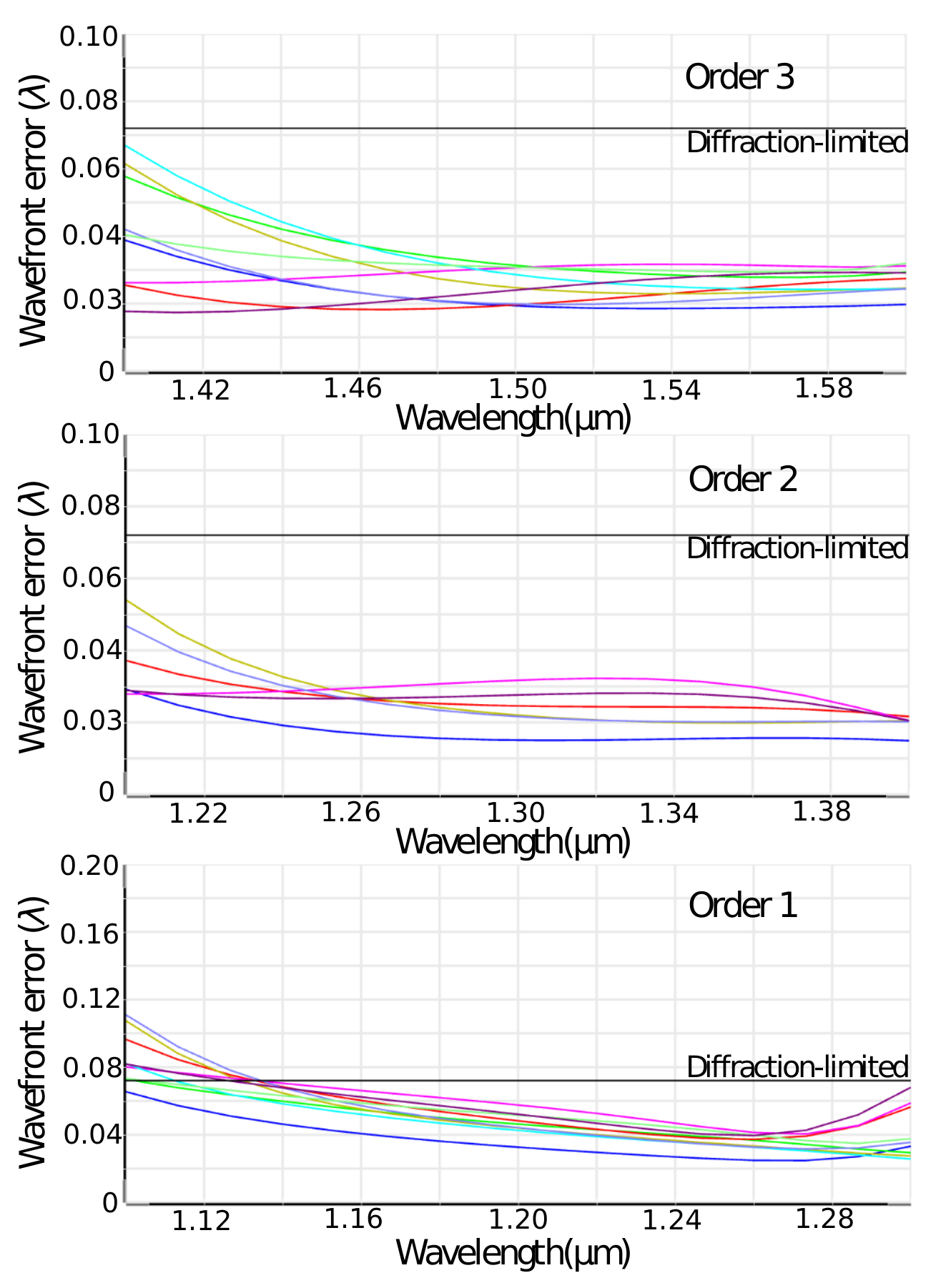}
	\end{center}
	\caption{The wavefront quality of the spectrograph as function of wavelength for the three different orders that are made by the \ac{VPHG}. The various colors represent a position along the pseudo-slit, and are the same pseudo-slit position for each order. Here we can see that the spectrograph design is diffraction-limited over almost the full spectral bandwidth. Only the blue part of the Y-band is slightly non diffraction-limited, which will lower the effective resolving power.}
	\label{fig:wavefront_rms}
\end{figure}

\begin{figure}
	\begin{center}
        \includegraphics[width=\textwidth]{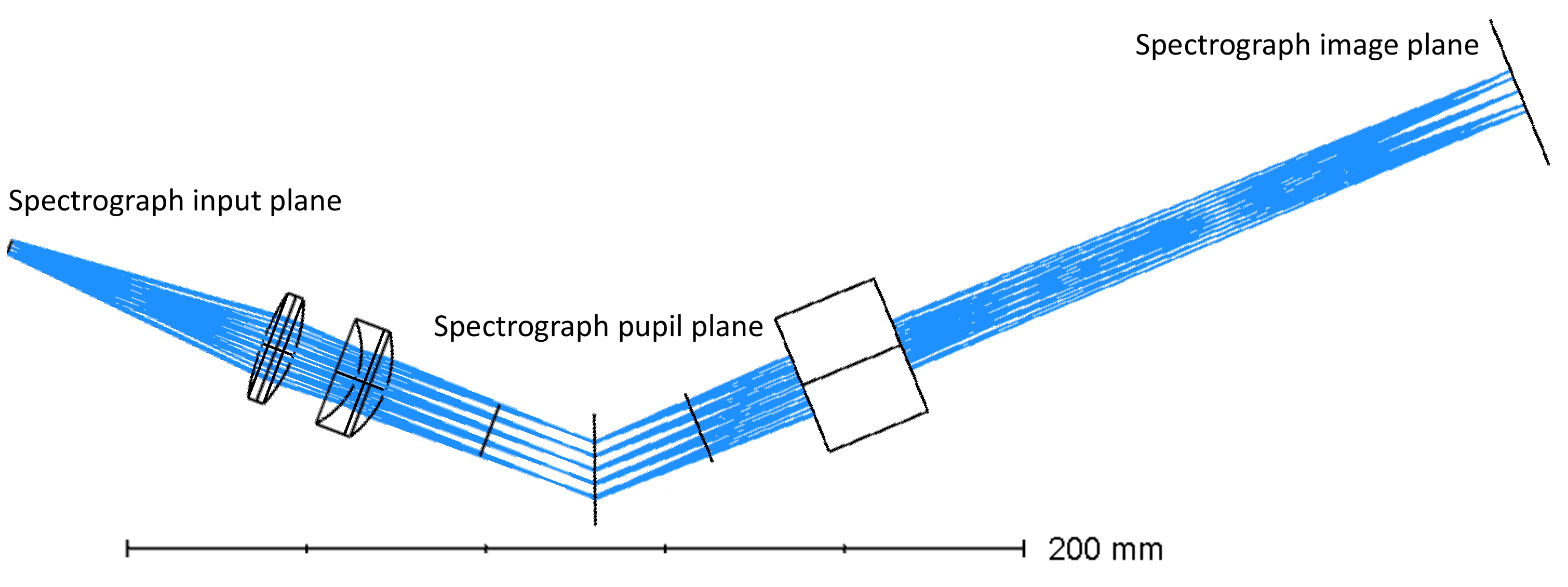}
	\end{center}
	\caption{The layout of the spectrograph with the different components of the spectrograph. The scale bar shows the small footprint of the optical path of the spectrograph, which is roughly 40\,cm by 20\,cm.}
	\label{fig:spectrograph_layout}
\end{figure}

\section{The spectrograph}
\subsection{Optical design}

The spectrograph was designed with off-the-shelf lenses. The output of the reformatter was estimated to have a focal ratio of 5, which was subsequently used as the input source for the spectrograph design. The spectrograph design is a standard first-order design, with a collimator lens that creates the beam that will be dispersed by the grating, and a camera lens that focuses the spectra onto a camera. For the camera lens we used an achromatic tube lens from Thorlabs, which has diffraction-limited performance over a large field of view and spectral bandwidth while also delivering a flat image plane. These properties makes the tube lenses from Thorlabs ideal as camera lenses for integral-field spectrographs. From the available tube lenses we found that the TTL200-S8 fitted within the requirements of the \ac{MCIFU}.

\textcolor{black}{The focal ratio of the waveguides (\#F=5) together with the effective slit length of 2\,mm required us to create a custom collimator.} We used combinations of spherical singlets to create a 'semi-custom' collimator. The design procedure started with two off-the-shelf doublets. The internal interface of each of the doublets was set to flat (a radius of curvature of infinite). All four outer surfaces of these two doublets were optimized by minimized the angular \ac{rms}. Then the surfaces were iteratively replaced with the closest matching off-the-shelf singlet that was available from either Thorlabs or Edmund Optics and then the angular \ac{rms} was minimized by re-optimizing the remaining surfaces. We repeated this procedure until we found a combination of singlets that had diffraction-limited performance over the full wavelength range and pseudo-slit length of the reformatter. While the optical quality of this semi-custom collimator was satisfying not all optics were available with \ac{NIR} optimized coatings. This resulted in a collimator with several uncoated optics that reduced the throughput. Figure \ref{fig:wavefront_rms} shows the total theoretical wavefront quality of the spectrograph design. Almost the full wavelength range is diffraction-limited, except for the small range of 1.0 -- 1.03\,$\upmu$m which is nearly diffraction-limited. The final layout of the \ac{MCIFU} spectrograph can be seen in Figure \ref{fig:spectrograph_layout}. The spectrograph could be kept small with a footprint of approximately 200\,$\times$\,400\,mm due to the \ac{SM} input.

The collimator creates an 11 mm pupil diameter, with \textcolor{black}{less than 0.5 mm of} pupil wander as a function of position along the pseudo-slit. The diffraction-limited resolving power of the spectrograph is $R=W\rho$, with $W$ the projected size of the pupil onto the grating and $\rho$ the line density of the grating. The theoretical maximum resolving power would then be approximately 8000, 6700 and 5800 for Y-band, J-band and H-band respectively. For best performance the spectrograph has been designed to work with an output plane of 2048x2048 pixels. A C-RED II with 640x480 pixels was available for the \ac{MCIFU} tests. The complete spectral range was covered by translating the camera. The full spectrograph output is covered by 12 camera positions.

\subsection{Mechanical design}

\begin{figure}
	\begin{center}
        \includegraphics[width=\textwidth]{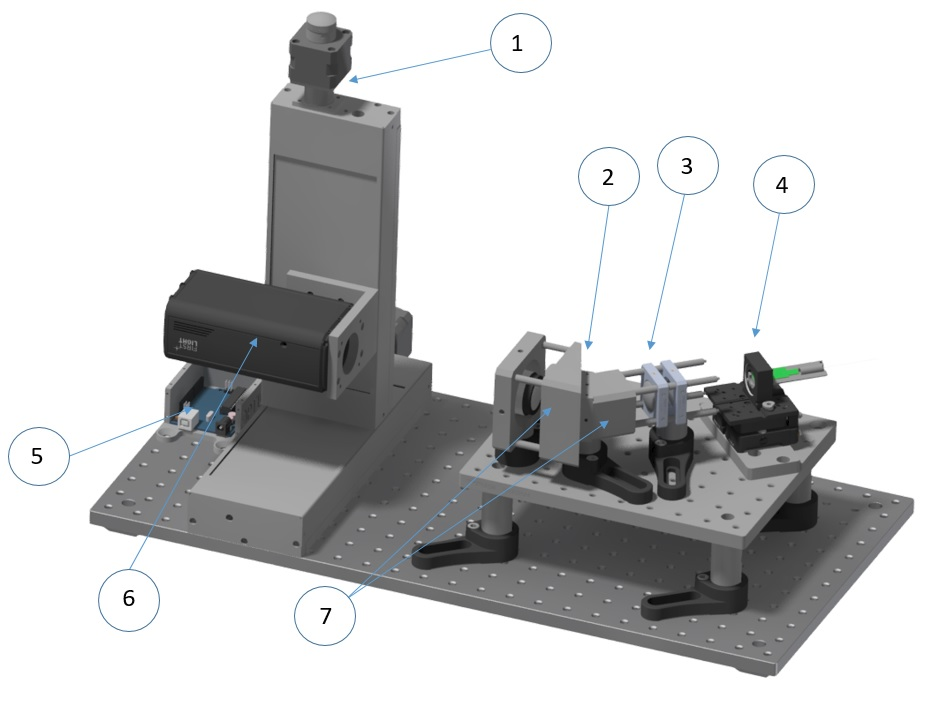}
	\end{center}
	\caption{Mechanical design of the spectrograph from the reformatter to the camera: 1) Motorized stages, 2) Dispersing element, 3) Collimating Optical elements, 4) Reformatter holder, 5) Arduino control components, 6) C-RED 2 camera, 7) 3D printed elements.
}
	\label{fig:mech_des}
\end{figure}

The mechanical design of the spectrograph ( see Figure \ref{fig:mech_des}) is simple thanks to its small size and the optical beam features (slow beams). Moreover, the simplicity was also a requirement due to budget and time constraints from the design to the commissioning. The working angle of the \ac{VPHG} in respect to the collimator and the angle with the camera are set using two 3D printed pieces, allowing for a simple integration and alignment procedure. The additive manufacturing approach was chosen also for the reformatter holder that allows for fitting the reformatted fiber inside a standard mounting. The degrees of freedom of this component are the tip-tilt, the de-center and the focus. The motorized stages that carry the camera are the same used in 3D printed setups and allows for covering the full field-of-view by a set of discrete positions.

\subsection{Performance}

After integration of the spectrograph we characterized the throughput and resolving power of the spectrograph. In order to have a good estimation of the spectral throughput of the instrument we used a Tunable Light Source (Oriel TLS-300X). The poor coupling between the faint TLS source and the fiber prevented a good signal/noise for the throughput measurement. In order to send some light through the instrument we had to measure subsystems independently and without the fiber. Moreover, as the $f/ \# $ of the source didn't match the instrument, we decided to first measure the Multiplexed \ac{VPHG}, and then the in-line optical train without the \ac{VPHG}.

\begin{figure}
	\begin{center}
          \includegraphics{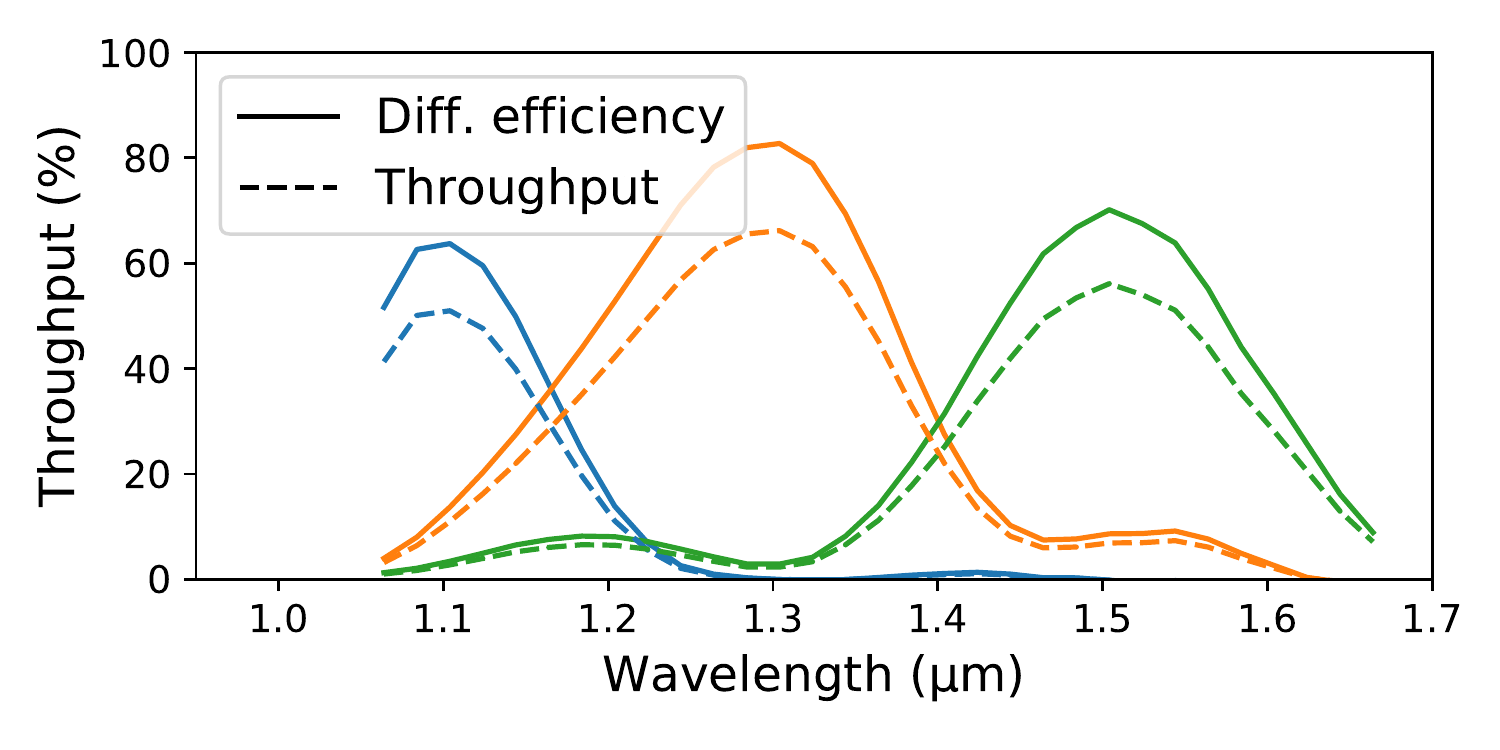}
	\end{center}
	\caption{The measured diffraction efficiency and throughput of the spectrograph without the fiber link as measured by the Tunable Laser Source. The solid lines show the diffraction efficiency of the stacked \ac{VPHG} system. The dashed lines include the throughput of the spectrograph optics. }
	\label{fig:meas_spec}
\end{figure}

The throughput of the spectrograph lenses was also measured with a 1064 nm diode laser, knowing that the spectral response is mainly due to interface reflections it is consistent with the previous estimations. The throughput of the lenses in the spectrograph is approximately 0.8 which is consistent with the fact that we the first two lenses are not AR coated. The total throughput of the system is then calculated adding the \acp{VPHG} contribution. In Figure \ref{fig:meas_spec} we report the spectrographs total throughput and the measured efficiency curves of the Multiplexed \ac{VPHG}.

\begin{figure}
	\begin{center}
          \includegraphics{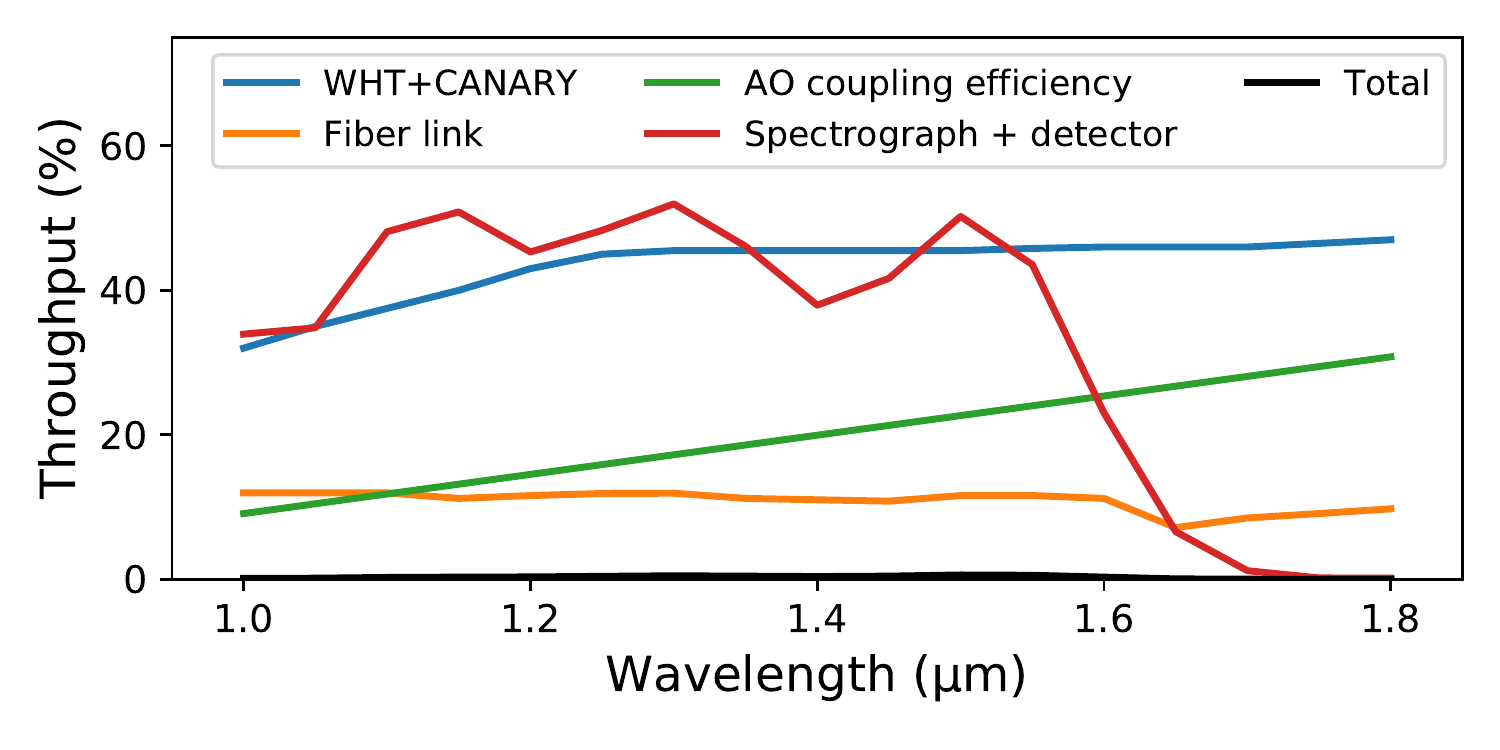}
	\end{center}
	\caption{Throughput of the individual components together with the total throughput (black). The largest limiting factor is the fiber link. The total throughput is 0.4\,\% on average. }
	\label{fig:e2e_throughput}
\end{figure}

\textcolor{black}{With all individual components measured we can determine the total end-to-end throughput. The wavelength dependent throughput can be seen in Figure \ref{fig:e2e_throughput}. The throughput of CANARY with the \ac{WHT} (\textit{T. Morris, private communication}) is also included in the throughput budget. The impact of the \ac{AO} performance on the coupling efficiency is based on the expected Strehl ratio of CANARY and simulations that determine the relation between the Strehl ratio and fiber coupling efficiency \cite{jovanovic2017injectsmf}. From this we determine that the expected end-to-end throughput is on the order of 0.4\,\%. The main limitations are the fiber link throughput and the AO performance.}

\begin{figure}
	\begin{center}
        \includegraphics{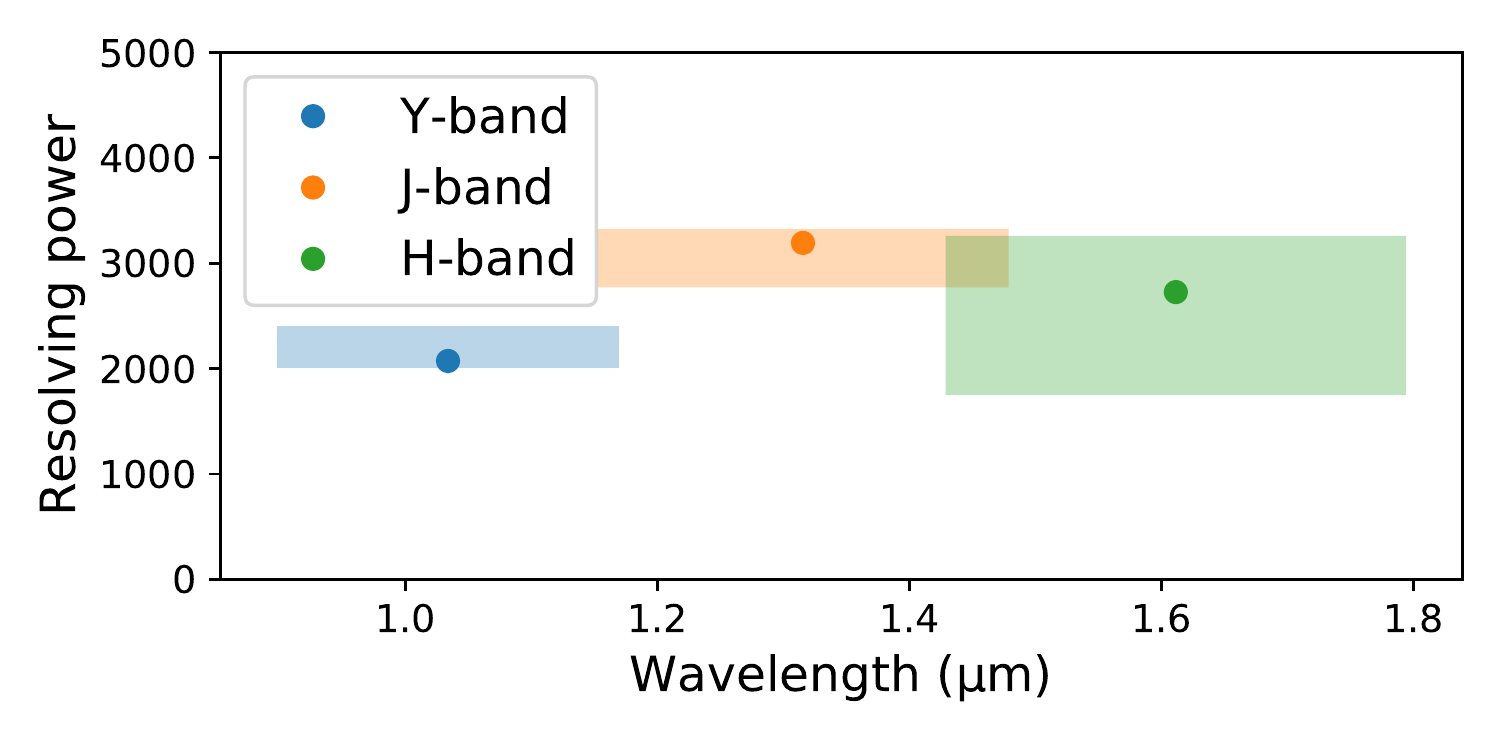}
	\end{center}
	\caption{The measured resolving power of the MCIFU for the three different spectral orders. The shaded area shows the 1-$\sigma$ resolving power estimates from several emission lines within each order. The points in each area show the median resolving power of that order. The H-band resolving power has a large spread due line blending of many of the measured lines. The estimated resolving power is a factor 2 to 3 lower than the theoretical limit.}
	\label{fig:resolvingpower}
\end{figure}

The resolving power of the spectrograph was determined with a Krypton line lamp (Oriel 6031 Kr Spectral Calibration Lamp).To obtain a wavelength solution for all individual cores we injected the light through the fiber bundle. Even though the line lamp was a large incoherent light source, we were still able to couple enough light into each individual core for calibration purposes. The wavelength solution was fitted with a first order polynomial. From the wavelength solution we derived a linear dispersion of 0.070\,\text{\AA}nm$^{-1}$, 0.850\,\text{\AA}nm$^{-1}$ and 0.095\,\text{\AA}nm$^{-1}$. The measured dispersion of each of the gratings was within 1\% of the designed dispersion, indicating that the triple stacked grating was manufactured within specifications. From the line lamp measurements we could also derive the effective resolving power of the spectrograph, which is defined as the center wavelength of the Kr emission lines divided by their \ac{FWHM}. The resolving power for the different spectral orders can be seen in Figure \ref{fig:resolvingpower}. The effective resolving power is a factor 2 to 3 lower than expected. This is most likely due to a misalignment of the collimation optics, as the \ac{PSF}s of each emission line \textcolor{black}{are magnified more than expected, show an asymmetry and have a strong Airy ring. These features are not expected from \ac{SM} waveguides that output Gaussian-like profiles, and indicate that we are vignetting the beam inside the spectrograph.}

\begin{figure}
	\begin{center}
        \includegraphics{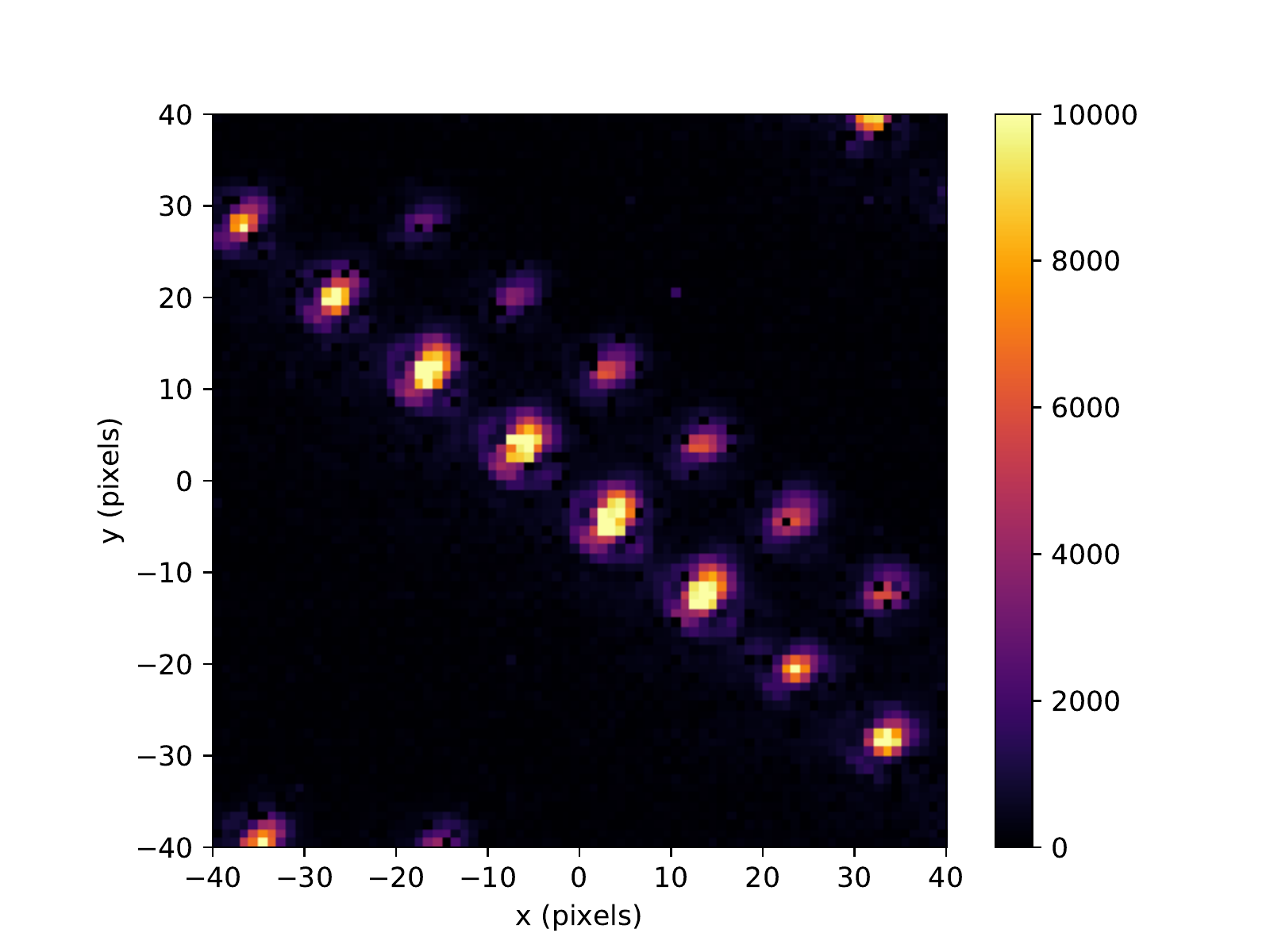}
	\end{center}
	\caption{The PSFs of the spectrograph as measured with the Kr spectral calibration lamp. The horizontal axis is the dispersion axis. The diagonal array of PSFs is the output of a single emission line illuminating a single row of the reformatter. The PSFs are extended and asymmetric towards the north-east and show a clear Airy ring.}
	\label{fig:psflets}
\end{figure}

\section{Testing with CANARY}

The spectrograph was tested with the CANARY \ac{AO} system at the \ac{WHT} \cite{gendron2016final}. CANARY is an \ac{AO} demonstration test bed for wide-field laserguide star tomography and open-loop \ac{AO} control. The main purpose of such an \ac{AO} system is to deliver good wavefront correction over a large field of view. CANARY was used in \ac{SCAO} mode for the \ac{MCIFU} experiment. In this mode the wavefront errors are measured by a $7\times7$ Shack-Hartmann wavefront sensor, which are then fed- back to a 52 element \ac{DM} \textcolor{black}{and dedicated tip-tilt mirror}. In this configuration CANARY delivers a Strehl of 30\,\% in nominal atmospheric conditions in H-band. The data was taken between the 18th and 21st July 2019, under the OPTICON open access time. This time was shared with two other groups performing experiments, dividing the nights into sections.

Once assembled, the \ac{MCIFU} was tested using the internal CANARY sources. We used a modified simulated annealing routine \cite{kirkpatrick1983optimization} to remove \ac{PA} between the wavefront sensor and the fiber. The algorithm optimised the shape of the \ac{DM}, maximising the fibre coupling. This increased the fiber coupling by approximately 10 \%.

\begin{figure}
	\begin{center}
        \includegraphics[width=\textwidth]{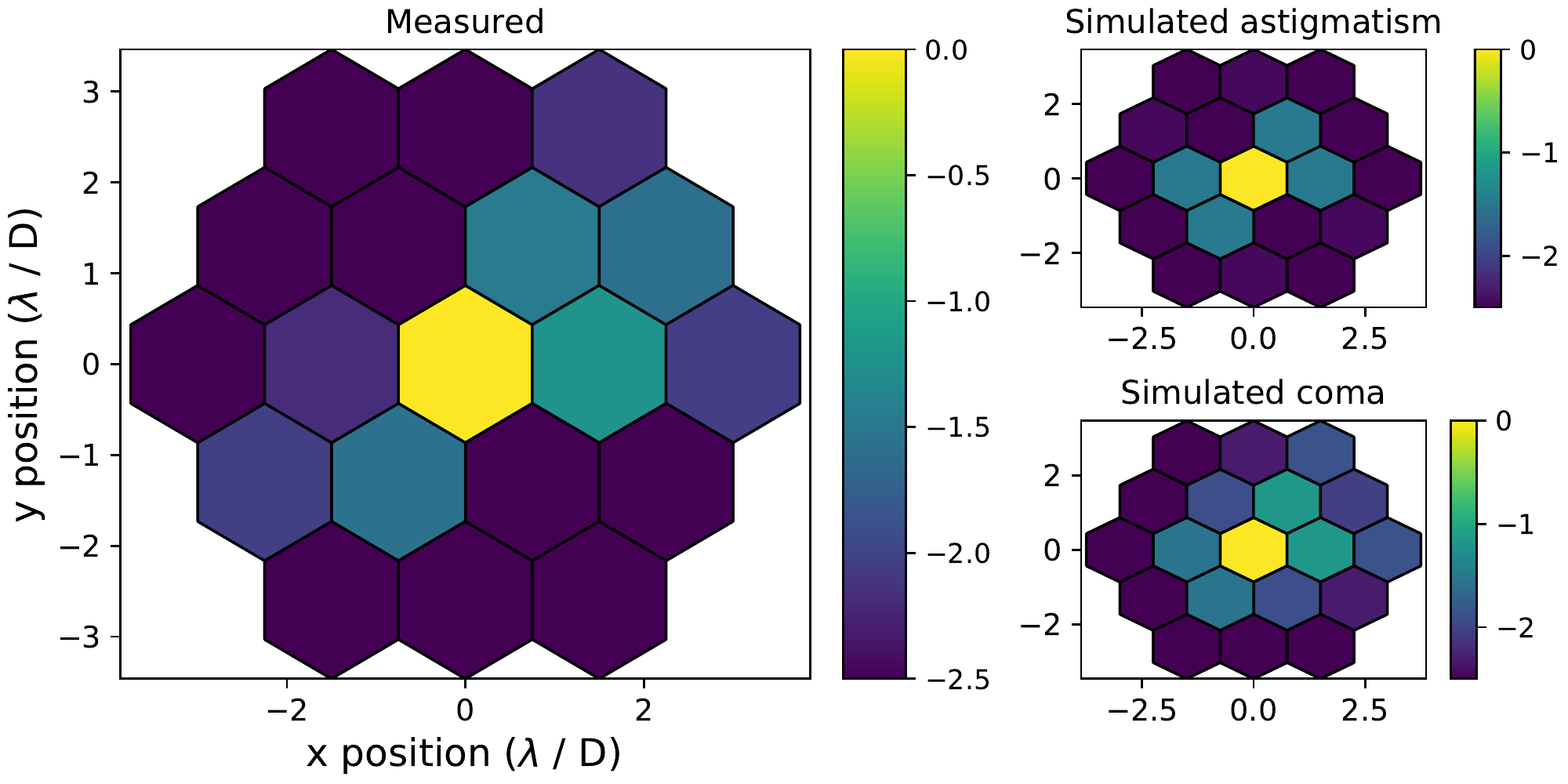}
	\end{center}
	\caption{The monochromatic post-fiber contrast map on a logarithmic scale as measured with a 1550nm laser (left). A comparison of the asymmetry in the illumination with simulations (right) hints on the presence of 0.5 radians rms residual astigmatism and/or coma.}
	\label{fig:laser_contrast}
\end{figure}

Several exposures with increasing exposure time were taken to create a high-dynamic range image for post-fiber contrast determination. The post-fiber contrast map can be seen in Figure \ref{fig:laser_contrast}. The contrast in the first ring is on the order of $10^{-2}$, which is similar to the contrast of the first Airy ring. The map also shows that there is some asymmetry in the \ac{PSF}. The asymmetry in the contrast map hints on residual astigmatism or coma\textcolor{black}{, which is on the order of 0.5 radians rms by comparing simulated contrast maps with the data. This can be seen in Figure \ref{fig:laser_contrast}.}

\subsection{On-sky performance}
\begin{figure}
	\begin{center}
        \includegraphics[width=\textwidth]{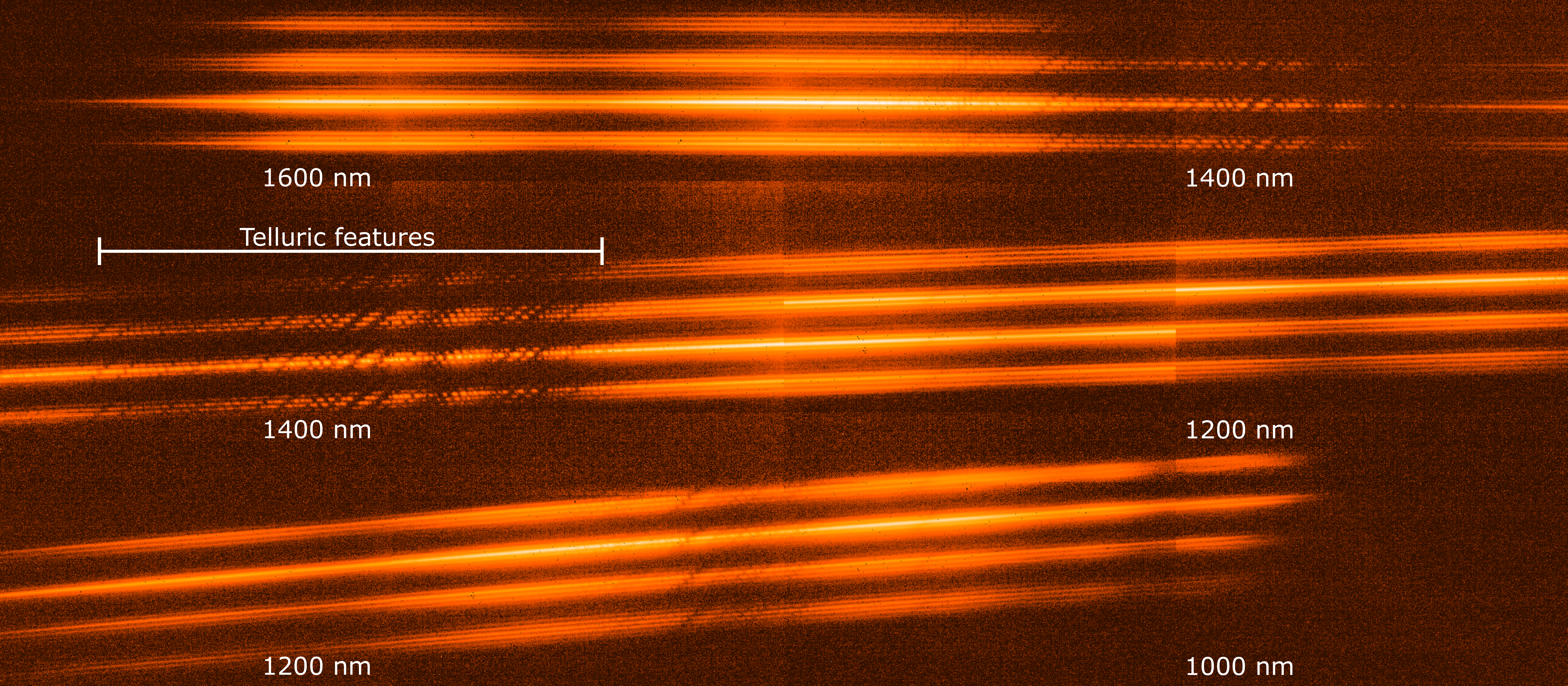}
	\end{center}
	\caption{The full spectrograph output of Vega after stitching 12 detector positions together. The color of the image is logarithmically stretched to highlight the spectra from the fainter fiber cores. Each fiber captures a different part of the PSF at the input, therefore the brightness of the spectral trace corresponds with the brightness of the captured part of the PSF. The main features that are visible are the telluric lines imprinted into the spectrum of Vega. Abrupt changes in the continuum of the individual cores are visible because the full image was reconstructed from 12 observations that were taken one after another, with varying conditions. The beginning and end of each spectrum has been marked with the corresponding wavelength.}
	\label{fig:vega_spectrum}
\end{figure}

Once on-sky we targeted bright stars, in order to estimate the performance. Our prime target was Vega, because we could use it as a calibrator star. In Figure \ref{fig:vega_spectrum} the full spectrograph output is shown by a stitched image of 12 camera positions.Due to the \ac{AO} tip-tilt variation and because the exposures of the individual camera positions were taken sequentially and the  there is some variability in the flux between the positions. Not all cores of the \ac{MCF} could be used during the on-sky demonstration as there was not enough separation between the three orders to fit all the 73 cores, which resulted in overlapped spectra. The top and bottom of the reformatter was blocked with a mask to remove the overlapping spectra.

\begin{figure}
	\begin{center}
       \includegraphics[width=\textwidth]{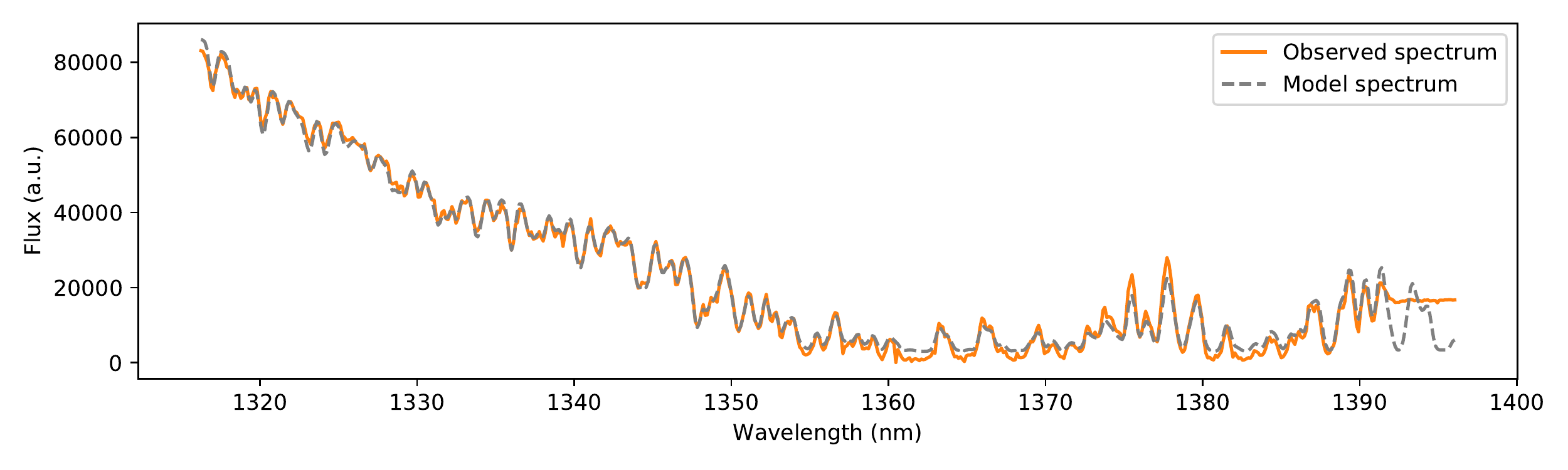}
	\end{center}
	\caption{The spatially averaged spectrum of Vega from 1300 nm to 1400 nm. The structure of this part of the spectrum of Vega is predominantly caused by the telluric absorption lines. }
	\label{fig:telluric_spectrum}
\end{figure}

Vega is an A0 star which means that it is almost featureless, the main spectral features that are visible come from the Earth's telluric absorption lines. Figure \ref{fig:telluric_spectrum} shows the extracted spectrum from 1310 nm to 1390 nm averaged over all fibers. We chose this part of the spectrum  as it illustrates the effect of the telluric features. The telluric features allowed us to do an independent measurement of the on-sky resolving power and wavelength solution. We used the ESO SkyCalc \cite{noll2012skycalc,noll2013moonlight} to generate a transmission spectrum using the standard Paranal atmospheric parameters. Vega was modelled with a high-resolution PHOENIX stellar model \cite{husser2013phoenix} with an effective stellar temperature of $T_{\rm{eff}}=9600$, metallicity of $Z=-0.5$, and surface gravity $\log g=4.0$. We used a 4th order polynomial for the instrument throughput and also used a 4th order polynomial for the wavelength solution. The high-resolution spectrum was convolved by a Gaussian with a certain width to mimic the effect of the spectrograph resolving power. The retrieved resolving power from this procedure was $R=2970$, which matches very well with the estimated resolving power from the Kr line lamp. We did not match all features in the observed spectrum because the parameters of both the telluric model and the stellar model were fixed. The good match between the measured and modelled spectrum shows that we can extract high fidelity spectra that are not contaminated by etalon and fringing effects. Previous \acp{SMF} spectrographs were plagued by strong etalon amplitude variations, that could be up to 10\,\% across the spectra\cite{rains2016rhea, rains2018rhea}.

Because of the optical layout of CANARY and the \ac{MCIFU} it was not possible to image the \ac{PSF} and the spectrograph output. Therefore we were not able to get a direct estimate for the on-sky fiber coupling efficiency.

\begin{figure}
	\begin{center}
        \includegraphics{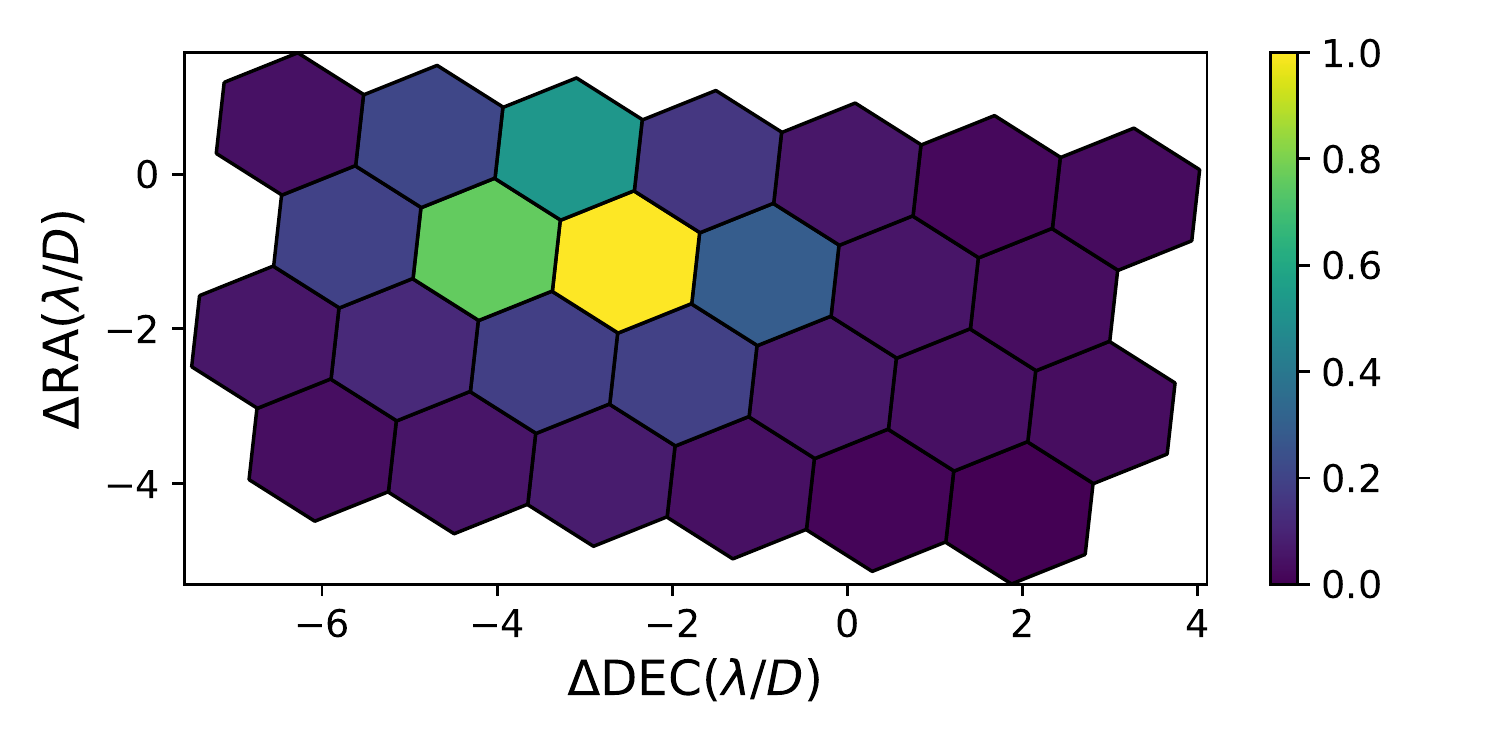}
	\end{center}
	\caption{The on-sky reconstructed H-band image, using the MCIFU, of the bright star Alpha Herculis. The observations were taken at 10Hz. The image shown here are is the mean of 1000 successive images from the data stream. Each micro-lens samples roughly 1.3 $\lambda/D$. The asymmetry in the PSF comes from a slow drift in tip/tilt alignment during the observations.}
	\label{fig:rasalgehi_collapsed_image}
\end{figure}

A second target that was observed was Alpha Herculis, one of the brightest infra-red targets in the sky with magnitudes of $J=-2.3$ and $H=-3.2$ \cite{cutri20032massphotometry}. The brightness of the target allowed us to do high-speed integral-field spectroscopy at 10 Hz \textcolor{black}{, at a single camera position.} In Figure \ref{fig:rasalgehi_collapsed_image} we show the reconstructed H-band \ac{PSF} of one of the exposures.

During the OPTICON July 2019 run at the \ac{WHT}, another experiment was performed \textcolor{black}{to characterise} \textcolor{black}{AO} performance \textcolor{black}{with a more sophisticated controller, the} \ac{LQG} regulator instead of the standard integrator. The \ac{LQG} regulator relies on a model that describes the temporal dynamics of the disturbance and whose relevance is a key for performance improvement. It includes both the auto-regressive order 2 model on the first 495 Zernike modes built from turbulence priors as described in \cite{sivo2014first}, and a data-driven model for the 9 first modes estimated from recent \ac{AO} telemetry data.  This \ac{LQG} controller provides a more accurate modeling of the temporal dynamics for the most energetic modes: not only vibrations on the tip and tilt but most of other sources of disturbance on the 9 first modes is captured whether it comes from dome turbulence, windshake or atmospheric turbulence. Implementing this \ac{LQG} controller required very little tuning and proved to deliver increased performance and to be much more stable than the integrator \cite{sinquin2020datacontrol}. During the night from July 21st to 22nd 2019, we observed Alpha Herculis both with the integrator and the \ac{LQG} and measured the impact on the tip-tilt residuals. To estimate performance, we made a white light image of the \ac{IFU} \ac{PSF} \textcolor{black}{and calculated the centroid of the white-light PSF as function of time. The centroid was measured by using the center-of-mass of each PSF.} We estimated that the \ac{LQG} successfully decreased the standard deviation of the \textcolor{black}{centroid} by a factor of approximately 1.4.

\section{Discussion}
\label{sec:Discussion}
The on-sky tests of the \ac{MCIFU} have shown that the novel components are able to work together on-sky to deliver a \ac{SMF}-fed \ac{NIR} \ac{IFS}. It was not possible to optimize the performance of all components due to the short time (6 months), between inception of the instrument and the on-sky demonstration. Because the \ac{MCIFU} is modular it is possible to redesign and optimize individual components and switch them for better performance. There are several parts that we have identified that can be improved with a clear path.

\subsection{Fiber link improvements}

Our complete fiber link system showed throughput of between a few to 27\,\% for individual cores. As this fiber link was a new system developed in only a few months we are happy with its initial performance. For future experiments we will develop a new fiber, with the aim of both increasing throughput and uniformity. Our target is to bring the fiber link throughput much closer to 50\,\%. 
To achieve the increase in the performance for our new reformatter, we will need to improve our ability to accurately control the inscription \textcolor{black}{in order to create a better mode matching between the cores and to optimize their alignment. A shorter pulse laser can allow us to optimize the laser-matter interaction in the substrate and enhance the inscription accuracy. By this way, we expect to increase the throughput for the straight waveguide and enable a more precise alignment of the MCF with the ULI wavguides to avoid the variations in throughput seen in Figure \ref{fig:HWfig2}.}

Further improvements in practical coupling efficiencies by the microlenses should also be possible as we keep developing the manufacturing process. Past work/experiences have shown that the coupling efficiency that can be achieved in the lab is close to the theoretical modeling. Yet, the \textcolor{black}{current} microlenses were \textcolor{black}{higher and} printed over a much larger surface than normal, likely effecting performance of the outer lenses. \textcolor{black}{The most probable cause is shrinkage of the structure during the curing step. Shrinkage has a stronger effect on the outer cores than the inner cores.} Thus further optimization and refinement/iterations need to be made to improve efficient coupling. \textcolor{black}{And finally an \ac{AR} coating can be added to the micro-lenses to get the highest possible throughput.}


\subsection{Spectrograph improvements}
The triple stacked \ac{VPHG} shows a performance close to the theoretical design limit. This indicates that we are operating at the current manufacturing limit of the \ac{VPHG} technology. The main limiting factor in the spectrograph is the collimation optics. The alignment of the separate singlets is quite sensitive and there is quite a large loss of throughput due to the uncoated optics. \textcolor{black}{Both the throughput and the resolving power} can improved by a redesign of the collimation optics.

\subsection{New detector}
Currently a C-RED 2 is used as a detector for the \ac{MCIFU}. The C-RED 2 is designed for a small field of view high-speed \ac{NIR} imaging and not long-integration spectroscopy. Due to the high read noise and dark current of the C-RED 2 the \textcolor{black}{single spectral channel limiting magnitude in 1 hour is about 12th magnitude at the \ac{WHT} if we exclude contributions from the background. Combining all spectral channels could improve this to about 15th magnitude.} Because of the low number of \textcolor{black}{detector pixels we need to scan the spectrograph focal plane in several steps, which as shown in this work makes post processing more difficult due to tip-tilt stability. Ideally we would use a detector with 2k $\times$ 2k pixels and low read and dark noise. Currently the best available solution for us would be a HAWAII-2RG \cite{kp2008hawki}, though we also note the increasing pixel numbers on SAPHIRA detectors and a 2k $\times$ 2k version would make a very attractive solution. Switching to a HAWAII-2RG would increase the limiting magnitude by \textcolor{black}{5} magnitudes.}

\subsection{On sky}
CANARY is an \ac{AO} demonstration test bed for wide-field laser guide star tomography and open-loop \ac{AO} control. For high-contrast imaging of exoplanets we need to use \ac{ExAO} systems that focus only on a very small field of view around the star. For \ac{SMF} optics this reduces the coupling into the fiber which is proportional to the Strehl \cite{jovanovic2017injectsmf}. For our observations, the estimated on-axis Strehl from CANARY was between 5-15\% in the H, which will be lower in the J band. \textcolor{black}{The Strehl was lower than expected due poor atmospheric conditions.} This shows an advantage of integral-field spectroscopy as opposed to single object spectroscopy where the full PSF is injected into a single SMF \cite{mawet2018kpic}. Due to the micro-lenses and multiple fiber the MCIFU is much less sensitive to pointing errors and residual tip/tilt errors. But adding the \ac{MCIFU} behind an \ac{ExAO} system that delivers high Strehl in the \ac{NIR} such as the \ac{MagAO-X} system \cite{males2018magaox} would allow for much higher coupling into the fiber. Due to the high Strehl it would also be possible to use high performance coronagraphs to reduce the amount of starlight that pollutes the planet signal. The past few years has shown that with \acp{SMF} it is possible to make higher performance coronagraphs \cite{mawet2017hdc, por2018SCARI, haffert2018SCARII,ruane2018vfn,coker2019mos}.

\textcolor{black}{We reckon that after implementing all proposed upgrades, the throughout can be significantly increased. An overview of the current and target throughput for each subsystem is shown in Table \ref{tab:throughput}. For the current system three components are significantly limiting the throughput, the AO performance, the micro-lens coupling and the reformatter. The throughput can be increased by a factor close to twenty if only these three components are addressed, which shows the importance of further refinement of the fiber link and future tests on ExAO systems.}

\begin{table}[ht!]
\caption{Wavelength averaged instrument throughput.}
\centering
 \begin{tabular}{||llll||} 
 \hline
 Sub-system & Current & Optimal & What to improve \\
 \hline\hline
  WHT+CANARY & 0.43 & 0.43 &  \\
 \hline
 Microlenses absorption & 0.90 & 0.96 & Thinner microlenses. \\
 \hline
 Microlens alignment & 0.3 & 0.9 &  More development time. \\
 \hline
 AO coupling & 0.2 & 0.66 & Switch to ExAO system with Strehl above 90\,\%. \\
 \hline
 Reformatter & 0.4 & 0.7 & Refinement of the manufacturing. \\
 \hline
 Spectrograph & 0.8 & 0.9 & New AR coatings. \\
 \hline
 VPH & 0.5 & 0.6 & Refinement of the manufacturing.\\
 \hline
 Detector & 0.8 & 0.9 & Switch to HAWAII-2RG. \\
 \hline
 Total & 0.003 (0.005) & 0.07 (0.11) & Peak throughput in brackets. \\
 \hline
\end{tabular}
\label{tab:throughput}
\end{table}

\section{Conclusions}
\acresetall

Here we present the \ac{MCIFU} an innovative prototype \ac{SMF}-fed \ac{IFS} designed to characterise exoplanets that can easily be added to existing and future high contrast imaging instruments. The \ac{PSF} from an \ac{AO} or \ac{ExAO} system is focused on a 3D printed \ac{MLA} which feeds a \ac{MCF}. At the entrance to the spectrograph the output slit is formed using a \ac{ULI} reformatter. The spectrograph itself is formed from off-the-shelf components, 3D printed components, a custom set of triple stacked \acp{VPHG} and a C-RED 2 detector. The prototype instrument itself went from design to telescope in approximately six months. Thanks to its modular nature the instrument can be easily upgraded and adapted for observations with other \ac{AO} systems.

The \ac{SMF}-link has been demonstrated to work on-sky, even in low Strehl conditions which are challenging for instruments where the full \ac{PSF} is injected into a single \ac{SMF}. We derived a resolving power consistent with our lab measurements from the telluric lines in the observed spectrum of Vega. And furthermore we demonstrated that it is possible to get high-speed integral-field spectra with the current \ac{MCIFU}. With our observations we could confirm the improved performance of a new \ac{AO} real time control algorithm.

Both the fiber link and spectrograph performance is slightly less than half the theoretical maximum, with the fibre link throughput being around 20\,\% and the resolving power being around 3000. This is due to the short manufacture and assembly time, and currently work is being done to improve and we plan to test the first modifications with MagAO-X. We will compare our results to those from CANARY and then begin a modular improvement program, bringing the system from prototype to full instrument.

\subsection*{Disclosures}
The authors declare no conflict of interest.

\subsection* {Acknowledgments}
\textcolor{black}{We want to thank the referees for giving extensive feedback which has greatly improved our work.} We thank B. Wehbe for sharing his knowledge of atmospheric dispersion and R. and D. Haynes for productive discussions about fibre preparation and packaging. We also thank the \ac{NYRIA} network, for seeding the ideas and beginning the collaboration for this project.

Sebastiaan Y. Haffert acknowledges funding from research program VICI 639.043.107, which is financed by The Netherlands Organisation for Scientific Research (NWO).

Support for this work was provided by NASA through the NASA Hubble
Fellowship grant \#HST-HF2-51436.001-A awarded by the Space Telescope
Science Institute, which is operated by the Association of Universities for Research in Astronomy, Incorporated, under NASA contract NAS5-26555. We also thank Covestro AG for providing samples of the Bayfol$^\text{\textregistered}$ HX materials.

Robert J. Harris is supported by the Deutsche Forschungsgemeinschaft (DFG) through project 326946494, `Novel Astronomical Instrumentation through photonic Reformatting'. B. Sinquin is funded by H2020 OPTICON No 730890.

This project has received funding from the European Union's Horizon 2020 research and innovation program under grant agreement No 694513 and No 730890, from the UK Science and Technology Facilities Council (STFC) -- STFC grant no. ST/N000544/1 and no. ST/N000625/1, from the Bundesministerium f\"ur Bildung und Forschung (BMBF), joint project PRIMA (13N14630), the Helmholtz International Research School for Teratronics (HIRST) and the Deutsche Forschungsgemeinschaft (DFG, German Research Foundation) under Germany's Excellence Strategy via the Excellence Cluster 3D Matter Made to Order (EXC2082/1 – 390761711).

\subsection* {Data, Materials, and Code Availability} 
\textcolor{black}{All data that was used to support this work can be found at 
\doi{10.25422/azu.data.12857885}.}


\bibliography{report}   
\bibliographystyle{spiejour}   


\vspace{2ex}\noindent\textbf{Sebastiaan Y. Haffert} is a NASA Hubble Postdoctoral Fellow at the University of Arizona's Steward Observatory. He received his PhD degree in astronomy cum laude from Leiden University in 2019. His research focuses on high-spatial and high-spectral resolution instrumentation for exoplanet characterization.

\vspace{2ex}\noindent\textbf{Robert J. Harris} completed his PhD at Durham University in 2014, modelling astrophotonic devices and spectrographs and developing photonic reformatters for high resolution spectroscopy. Following the PhD and a STEP postdoc (again at Durham), he took up a Carl-Zeiss fellowship at the Landessternwarte, Heidelberg, where he was also awarded the Gliese fellowship. He is currently working at the MPIA in Heidelberg.

\vspace{1ex}
\noindent Biographies and photographs of the other authors are not available.

\listoffigures
\listoftables

\end{spacing}
\end{document}